\documentclass[a4paper,fleqn,usenatbib]{mnras}

\usepackage{newtxtext,newtxmath}
\usepackage[T1]{fontenc}
\usepackage{ae,aecompl}
\usepackage{booktabs,caption}
\usepackage[flushleft]{threeparttable}

\usepackage[nolist,nohyperlinks,printonlyused]{acronym}

\usepackage{float}

\usepackage{graphicx}	% Including figure files
\usepackage{amsmath}	% Advanced maths commands
\usepackage{comment}

\newcommand{\HI}{\ion{H}{I}}

\newcommand{\mathHI}{{\mbox{\scriptsize \HI}}}

\newcommand\given[1][]{\:#1\vert\:}

\newcommand{\lyaf}{\text{Ly$\alpha$ forest}}
\newcommand{\lya}{\text{Ly$\alpha$}}

\newcommand{\NHI}{$N_{\mathHI{}}$}

\newcommand{\NHIt}{N_{\mathHI{}} }

\newcommand{\vpfit}{\texttt{VPFIT}}
\newcommand{\bndist}{$b$-$N_{\mathHI{}}$~distribution}
\newcommand{\bn}{$\left\{b, N_{\mathHI{}}\right\}$}
\newcommand{\dndz}{d$N$/d$z$}

\defcitealias{Danforth2016}{D16}

\title[The IGM Thermal and Ionization State at $z < 0.5$]{A Measurement of the Thermal and Ionization State of the IGM at $z < 0.5$}

\author[Hu et al.]{Teng Hu,$^{1,2}$\thanks{E-mail: tenghu@ucsb.edu (UCSB)}
Vikram Khaire$^{3}$,
Joseph F. Hennawi$^{2,4}$,
Todd M. Tripp$^{5}$,
Jose O\~norbe$^{6}$, \newauthor
Michael Walther$^{7,8}$, and
Zarija Luki\'c$^{9}$ 
\\
$^{1}$Aix Marseille Univ, CNRS, CNES, LAM, Marseille, 13013, France\\
$^{2}$Physics Department, Broida Hall, University of California Santa Barbara, Santa Barbara, CA
93106-9530, USA\\
$^{3}$Department of Physics, Indian Institute of Technology Tirupati, Tirupati, Andhra Pradesh 517619, India\\
$^{4}$Leiden Observatory, Leiden University, PO Box 9513, NL-2300 RA Leiden, the Netherlands\\
$^{5}$Department of Astronomy, University of Massachusetts, Amherst, MA 01003, USA\\
$^{6}$Facultad de F\'isica, Universidad de Sevilla, Avda. Reina Mercedes s/n, Campus de Reina Mercedes, E-41012 Sevilla, Spain\\
$^{7}$University Observatory, Faculty of Physics, Ludwig-Maximilians-Universität München, Scheinerstr. 1, 81679 Munich, Germany\\
$^{8}$Excellence Cluster ORIGINS, Boltzmannstr. 2, 85748 Garching, Germany\\
$^{9}$Lawrence Berkeley National Laboratory, Berkeley, CA 94720, USA\\
}

\date{Accepted XXX. Received YYY; in original form ZZZ}

\pubyear{2026}

\begin{document}
\label{firstpage}
\pagerange{\pageref{firstpage}--\pageref{lastpage}}
\maketitle

\begin{abstract}

We apply a machine-learning-based inference method 
that exploits the joint Doppler parameter–column density ($b$–$N_{\mathHI{}}$) distribution from \lyaf{} decomposition to measure the thermal and ionization state of the intergalactic medium (IGM) in four redshift bins spanning $z=0.06$ to 0.48, using 82 archival 
% HST COS %% VK2 expanded below 
quasar spectra from the Cosmic Origin Spectrograph (COS) on board Hubble Space Telescope (HST). Our results show that the low-$z$ IGM ($z<0.5$) is extremely hot and nearly isothermal, 
%% VK2 I have modified the text below to address referee's comment
with $\log(T_0/\mathrm{K}) = 4.45^{+0.08}_{-0.12}$
($T_0 = 28183^{+5700}_{-6804}\,\mathrm{K}$) and
$\gamma = 1.06^{+0.13}_{-0.09}$ at $z=0.1$.
This temperature lies $\approx 7\sigma$ (and $7$ times) above the canonical prediction
($\log T_0 \approx 3.60$, i.e.\ $T_0 \sim 4000\,\mathrm{K}$, with
$\gamma \sim 1.6$ at $z=0$), where the IGM is expected to have cooled
long after He\,\textsc{ii} reionization.
%% VK2 note that the sigma is calculated using  the log-space number since that is where we report the posterior
%$T_0 = {28183}^{+5700}_{-6804}$ K and $\gamma = {1.06}^{+0.13}_{-0.09}$ at $z=0.1$, which is about seven times hotter than the canonical theoretical model, where the IGM is predicted to cool down at low-z ($T_0 \sim 4000$ K, $\gamma \sim 1.6$ at $z=0$), long after the conclusion of HeII reionization.
We also measure the hydrogen photoionization rate to be $\log (\Gamma_{\mathHI}/\text{s}^{-1}) = {-13.70}^{+0.10}_{-0.08}$ at $z=0.1$, which is about 
%60\% lower than the theoretical model 
$\approx 4\sigma$ below the range predicted by current UV-background synthesis models ($ \approx -13.3$). %% VK2 change
To investigate the discrepancy between these high temperatures and theoretical models, we assess the impact of small-scale turbulence. By exploring a parameter grid in turbulent velocity ($v_{\text{tur}}$) and $\Gamma_{\text{HI}}$, we find that a standard IGM thermal and ionization state combined with unresolved turbulence of $v_{\text{tur}} \simeq 15$ km s$^{-1}$ can successfully reproduce the observed line widths at $z=0.1$. Comparisons with high-resolution Space Telescope Imaging Spectrograph (STIS) %% VK2 expanded 
data indicate that the observed line widths are unlikely to be caused by instrumental resolution effects. Our findings suggest that either new heating mechanisms or unresolved turbulence are required to explain the unexpectedly broad Ly$\alpha$ lines observed in the low-$z$ IGM.

\end{abstract}


%%%%%%%%%%%%%%%%%%%%%%%%%%%%%%%%%%%%%%%%%%%%%%%%%%

%%%%%%%%%%%%%%%%% BODY OF PAPER %%%%%%%%%%%%%%%%%%

\section{Introduction}
\label{sec:intro}

The \ac{IGM}, being the largest reservoir of baryons in the Universe, plays a fundamental role in the formation and evolution of cosmic structures. 
After H~{\sc i} reionization \citep[$z \lesssim 6$, e.g.,][]{Becker2001,Fan2006,mcgreer1}, and He~{\sc ii} reionization which is completed by around $z\approx3$ \citep[see e.g.,][]{Shull2010, Worseck2011, Khaire2017},
% VK I would just keep the observational references directly related to HI reionization. Corrected by removing a few and adding a few.  Original is commented below. 
%\citep{Madau1998,Fan2006,Faucher-Giguere2008,Robertson2015,mcgreer1}, 
the thermal state of the IGM is mainly determined by the balance between heating from photoionization by the extragalactic UV background (UVB) and cooling
via Hubble expansion, radiative recombinations, and inverse Compton scattering of electrons off of the cosmic microwave background.
As a result of these processes, after the epoch of H~{\sc i} reionization, the IGM subsequently adheres to a power-law temperature-density ($T$-$\Delta$) relation: 
\begin{equation}
T(\Delta) = T_0 \Delta^{ \gamma -1},
\label{eqn:rho_T}
\end{equation}
where $\Delta = \rho/ \bar{\rho}$ is the overdensity, $T_0$ is the temperature at mean density $\bar{\rho}$,
and $\gamma$ is the power-law index \citep{hui1, McQuinn2016}. 
These two parameters $T_0$ and $\gamma$, thus characterize the thermal state of the IGM, 
and enable us to impose constraints on its thermal history at various epochs
\citep{schaye1999, mcdonald2001, ricotti2000, Dave&Tripp2001}
%\citep{Dave&Tripp2001, Becker2011, Rorai2017, Hiss2018, Gaikwad2021},
% VK Just cite old references here. Others are anyway cited later
% VK I have corrected it
%This helps us understand the thermal evolution of the IGM and
and its relationship to fundamental heating and cooling processes. 
%which enhance our understanding of the IGM thermal evolution and illustrate the intrinsic heating and cooling mechanisms of the Universe.

At high redshifts ($z>2$), current theoretical models of the IGM, particularly regarding its thermal and ionization states, align remarkably well with high-resolution Ly$\alpha$ forest observations \citep[e.g.][]{Bolton&Haehnelt2007, Becker2011, Becker&Bolton2013, Hiss2018, WaltherM2019, Gaikwad2021}. 
This concordance has established confidence in these IGM models, thereby leading to studies using the \lyaf~for probing cosmology \citep[e.g.][]{mcdonald2001, Busca2013}, measuring neutrino masses \citep[e.g.][]{McDonald2006, yeche2017, Garny2012}, and testing alternate models of dark matter \citep[e.g.][]{Viel2013, Armengaud2017, PD2020, Irsic2024}.
However, in the past decade, low-$z$ ($z<0.5$) observations of the IGM, facilitated by the \ac{COS} on board the \ac{HST}
% VK added acronym
% VK modified text commented below
%have shown discrepancies that challenge our current understanding of the IGM. These discrepancies signal an intriguing issue with the otherwise stellar concordance between theory and observations of the IGM.
have revealed discrepancies that challenge our current understanding, highlighting issues with the otherwise 
%stellar concordance
excellent agreement %% VK2 change
between theory and observations of the IGM at high-$z$.

A major unresolved issue in the low-$z$ IGM is the discrepancy between the observed and simulated distributions of the Doppler widths ($b$-parameter) of the Ly$\alpha$ absorption lines \citep{Gaikwad2017, Viel2017}. 
Observations show that the distribution of the $b$-parameter at $z < 0.5$ is broader and shifted to values notabley higher by $\sim$ 10 km/s than those predicted by simulations. 
Such higher-than-expected line widths suggest either a significant source of additional turbulence in the IGM of the order $\sim 10$ km/s, which is not captured by current simulations \citep[e.g.][]{Nasir2017, Bolton2022}, or that the low-$z$ IGM is substantially hotter than expected \citep[e.g.][]{Viel2017}.
% VK: adding below the Dark matter reference
Interestingly, one of the proposed resolutions of this discrepancy involves non-standard heating of the IGM via dark photon dark matter \citep{Bolton2022DarkPhoton}.

In fact, one of the fundamental predictions of the theory of the IGM is that,
a few hundred million years after the completion of He~{\sc ii} reionization,
which is supposed to be completed around $z\approx3$ 
\citep[see e.g.,][]{Shull2010, Worseck2011, Khaire2017},
the IGM should cool down primarily because of the adiabatic cooling caused by Hubble expansion. 
The IGM quickly loses memory of the thermal impact of He~{\sc ii} reionization, and asymptotes towards a temperature $T_0 \simeq 4000$ K completely determined by the shape of the UVB spectrum \citep[][]{McQuinn2016,Bolton2022}. 
Yet,  this anticipated cooling of the IGM at $z<1$ remains unverified by empirical data, leaving a 10-billion-year stretch of cosmic time uncharted in terms of the IGM thermal history.

Part of the challenge lies in the fact that for $z \lesssim 1.7$, the \lya{} transition is below the atmospheric cutoff ($\lambda \sim 3300$ \AA{}), demanding UV observations from space via HST, 
the only space-based telescope that has FUV and NUV spectrographs capable of observing Ly$\alpha$ forest required for these measurements. 
Currently, the only measurements of the IGM thermal state for $z < 1.5$ comes from  \citet{ricotti2000} and  \citet{Dave&Tripp2001} at $z=0.1$ and \citet{Hu2023b} at $0.9<z<1.5$, whereas the former are derived from datasets with very limited size ($\sim$ 50 \lya{} absorption lines), causing a substantial error margin, where $T_0 \sim 5000$ K with $\sigma_{T_0}$ > $5000$ K  at $z =0.1$.
Meanwhile, although \citet{Hu2023b} suggest that the IGM does not cool as expected at $z<1$, where $T_0 = 14100$ with $\sigma_{T_0} = $ 3200K at $z =1.0$, the measurements are restricted to $0.9<z<1.5$, leaving significant gaps in our understanding of the low-redshift IGM. Additionally, the uncertainties due to the limited data size implies that our knowledge of the IGM thermal state at low-$z$ remains imprecise.

At high-$z$, various statistics of \lyaf{} have been used to measure $T_0$ and $\gamma$ 
\citep[at $z>1.7$ e.g][]{Lidz2010, Bolton2010,Garzilli2012,bolton2014,Rorai2018, Gaikwad2021}. 
At low-$z$, in addition to the number of spectra, the sparsity of the \lyaf{} caused by low opacity further reduces the number of available \lyaf{} data.
%% JFH2 This setnence does not make sense to me. I'm not sure what you are trying to say. The opacity is what makes the b-N distributino possible at low-z but challenging at high-z. I don't follow your point about available data regarding opacity. 
%%TH I mean the opacity is very lower with IGM at z > 1.7, where data are easier to obtain.
To overcome this problem, \citet{Hu2022} adopted a novel inference method to jointly measure the IGM thermal state [$T_0$, $\gamma$] and its ionization state quantified by the H~{\sc i} photoionization rates, $\Gamma_{\mathHI{}}$ based on the decomposition of the \lya{} forest into $b$ parameter and column density $N_{\mathHI{}}$.
Under this method, Bayesian inference 
% VK was it Bayesian or just max. likelihood? 
of the model parameters 
[$\log T_0$, $\gamma$, $\log \Gamma_{\mathHI{}}$] is performed using the
joint 2D \bndist{} and the line abundance \dndz{}.
To facilitate this, we used neural density estimators \citep{Alsing2019} 
% VK add for emulating 2D b-N dist
and Gaussian emulators \citep{Ambikasaran2016},
% VK add for emulating dN/dz
both trained on a suite of Nyx 
%% Nyx hydrodynamical simualtions. Add the %% Lukic 2015 reference here as well. %%TH done
simulations \citep{nyx,Lukic2015}
consisting of 51 simulation models with different thermal histories \citep[presented in][]{Walther2017,Hiss2018}. 
This method takes into account the crucial degeneracy between thermal state and $\Gamma_{\mathHI{}}$, providing precise measurements of all three parameters $T_0$, $\gamma$, and $\Gamma_{\mathHI{}}$ even with a small dataset.

Using this method, we measured the thermal and ionization state of the IGM at $0.9<z<1.5$ from twelve \ac{HST} \ac{STIS} quasar spectra, hinting that the IGM does not seem to follow the expected cooling at $z<1$ \citep{Hu2023b}.
These measurements and the larger-than-expected $b$-parameters at $z < 0.5$ both point in the same direction -- the low-$z$ IGM appears to be much hotter (or more turbulent) than expected.
%, and there is no evidence that it is cooling down. 
Understanding this could require non-standard IGM heating mechanisms such as $\gamma$-ray 
blazars \citep{Puchwein2012}, dark matter annihilations \citep{Ripamonti2007, Araya2014}, dust heating \citep{Inoue2010,Bolton2022}, 
%% JFH What about the dust heating? You should mention that. 
%%TH: revised
or exotic dark matter models like dark photons \citep{Bolton2022DarkPhoton}. 
%whereby resonant conversion of dark matter to low-frequency photons can heat the IGM 
Alternatively, it could be, contrary to theoretical expectations, that some physical process, possibly powered by galaxy formation feedback, can
drive turbulent motions in the IGM that are not yet captured by simulations
\citep[see also][]{Viel2017, Hu2023}.

%% JFH2. I don't think is ithe precisoin of the dataset. It is the redshit coverage!! 
%%TH revised
However, due to the limited sample size, the precision of the measurements in \citet{Hu2023b} is insufficient to confirm the expected cooling of the IGM. More precise measurements of the IGM thermal state, particularly at $z<0.5$, are crucial in determining if the IGM deviates from theoretical expectations. Motivated by this, we present the IGM thermal state measurement at $z<0.5$ in this paper. 

%% JFH2 Better transition here. In this paper, in this paper appears twice. 
In this work, we employ the aforementioned method used in \citet{Hu2023b} to precisely measure both the thermal and ionization state of the IGM in four redshift bins from $z=0.06$ to $0.48$. We make use of 82 archival \ac{HST} \ac{COS} quasar spectra with high \ac{SNR}s. We make use of the corresponding metal identification from the low-$z$ IGM survey published by \citet[][hereafter \citetalias{Danforth2016}]{Danforth2016}. These spectra are fitted to obtain our \bn{} sample using a standard Voigt profile fitting code, \vpfit{} (see \S~\ref{sec:vpfit_z01}).  We then measure the thermal and ionization state of the IGM in the four redshift bins centred at $z=0.1$, 0.2, 0.3 and 0.4 
%describe the choice of the z-bins in chapter2
following the method described in \citet{Hu2022}.
In addition, we also explore alternative explanations for the observed discrepancy in the distribution of the $b$ parameter, such as small-scale turbulence missed by simulations and the potential overestimation of the HST COS resolution.

This paper is organized as follows. Section \ref{sec:data_z01} describes our observational data and the data processing methods, including continuum fitting, Voigt profile fitting, and metal masking. Section \ref{sec:simulations} details our hydrodynamic simulations, the parameter grid, and the methods for generating mock data, such as producing the Ly$\alpha$ forest from simulations, creating mock sightlines, and forward-modeling. In Section \ref{sec:inference}, we present our inference algorithm, including the use of emulators and the formulation of the likelihood function. We present the inference results in Section \ref{sec:result_z01}. We then discuss these results and explore alternative interpretations in Section \ref{sec:discussion_z01}. Finally, we summarize the key findings of this study in Section \ref{sec:conclusion_z01}.
%Throughout this paper, we use $\log$ to denote $\log_{10}$. 
The cosmological parameters used in this study ($\Omega_m = 0.319181, \Omega_b h^2 = 0.022312, h = 0.670386, n_s = 0.96, \sigma_8 = 0.8288$) are taken from \citet{Planck2014}.

\section{Observational Data}
\label{sec:data_z01}

The dataset we analyze is the publicly available compilation\footnote{\url{http://archive.stsci.edu/prepds/igm/}} of high signal-to-noise ratio (SNR) \ac{HST}/\ac{COS} 
spectra published by \citet{Danforth2016}. Consisting of 82 quasar spectra observed between 2009 and 2013 with the G130M ($1135 \sim 1450$ \AA{}) and G160M ($1360 \sim1775$ \AA{} ) gratings, 
%% JFH Can you be more specific and state how many spectra or how much pathelength in each setup that you have. 
%%TH done
this dataset represents the largest publicly available low-$z$ survey of the \lyaf{} to date, with total \lya{} path length $\Delta_z =$ 4.43. 
The nominal resolution of \ac{COS} is $R\sim 15000-20000$ depending on the wavelength and grating, which corresponds to roughly $15 - 20$ km $\text{s}^{-1}$,
and has a non-Gaussian \ac{LSF}.\footnote{While the COS LSF is close to Gaussian, it shows very broad wings. For more reference, see \url{https://hst-docs.stsci.edu/display/COSIHB/3.3+The+COS+Line-Spread+Function}} 
Individual spectra were co-added, taking into account all exposures and gratings, and then continuum-fitted by \citet{Danforth2016}.

\begin{figure*}
\centering
    \includegraphics[width=0.95\linewidth]{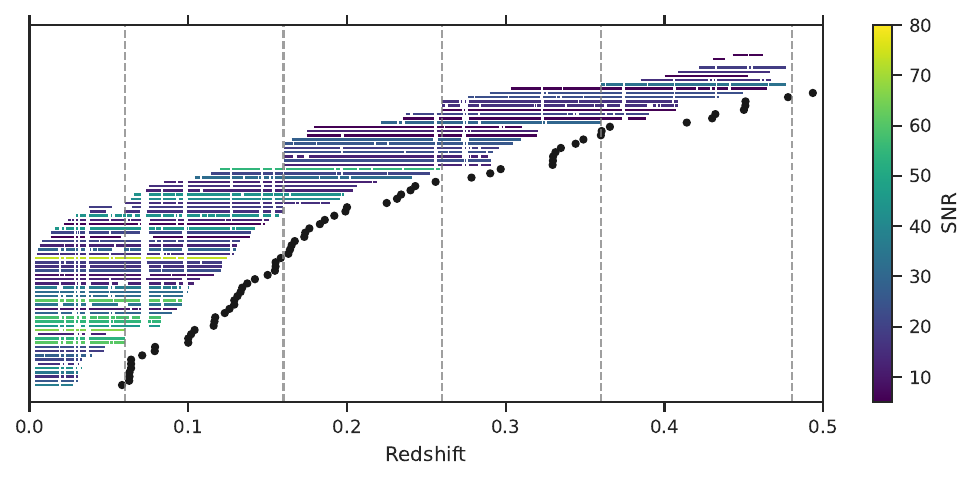}
  \caption{ The HST COS spectra used in this study. The quasar are shown as black dots,
  and the \lya{} spectra, with proximity zones removed, are shown as line segments. The colour indicates the mean SNR (per pixel) of a spectrum, and the gaps represent the masked regions.
  The four redshift bins used in this study are shown by the vertical dashed lines. The dots that represent the redshift for the few right-most QSOs are missing in the plot, since their redshift is above 0.5.}
  \label{fig:data_coverage_D16}
\end{figure*}

In this study, we focus exclusively on the \lyaf{}. Therefore, we utilize only the \lya{} regions, omitting Ly$\beta$ and higher Lyman series absorption lines at wavelengths shorter than 1050~\AA, and masking the quasar proximity zones at wavelengths greater than 1180~\AA~(see Fig.\ref{fig:data_coverage_D16}). This limits our analysis to the spectral segment with rest frame wavelengths between 1050~\AA{} and 1180~\AA{}. 
%% VK2 adding the text below for justification of not using z<0.06
This selection provides continuous Ly$\alpha$ forest coverage over
$0.06 < z < 0.48$, which is the redshift range we analyse in this work.
We do not use the lowest-redshift portion of the data ($z < 0.06$): the
$z \approx 0$ Ly$\alpha$ forest falls in the bluest part of the COS
spectra, near the Galactic Ly$\alpha$ line at $1215.7$~\AA, where it is
contaminated by the damped Milky-Way Ly$\alpha$ absorption together with
geocoronal Ly$\alpha$ and \ion{O}{i} airglow emission.
To mitigate edge effects at the spectral edges, quasar sightlines are segmented and padded with white noise based on the noise vector of the spectrum before being fed into the VP-fitting program. These padded regions are subsequently masked in post-processing. We apply similar edge treatments to the mock forward models to maintain consistency in our analysis.

%% JFH what about spectra observed with 
%% mulitple setups. Are those excluded or how do you deal with those?
%TH I think Danforth co-added the overlapping regions for different gratings. But the affected region is not huge. What I did id cope the the spectra at 1410 for both OB data and sim, adn fit <1410 with G130M and fir >1410 with G160M.
We use our automated Voigt profile (VP) fitting program, which is described in \S~\ref{sec:vpfit_z01}, to identify and fit all absorption lines in the 82 spectra to obtain the \bn{} datasets at each redshift bins. To ensure that our \bn{} datasets, obtained after VP-fitting, contains only \lyaf{} absorption lines, we mask all metal absorbers identified by \citet{Danforth2016}, which include both intervening metal lines and the lines arising from the \ac{MW} absorbers at $z=0$ and geocoronal emission lines. 
%% JFH not sure I follow how you are treating emission lines? Maybe you mean geocoronal, but you should state that explicitly.
%%TH revised
We also mask all Lyman series lines that are not \lya{}. All masks are then adjusted by eye to include the full emission profiles and gaps in the wavelength coverage. The full procedures for generating masks are detailed in \S\ref{sec:metal_ID_z01}.
%% JFH Here I would say that in the metal-id process Ly-beta or Ly-series lines may have nevertheless been used, i.e. to check whether a line is hydrogen or somethiung else, i.e. by Danforth. 
%%TH yes they do check Ly-beta or Ly-series lines, I have revised the text accordingly.

We show the redshift path covered by the data segments after masking in Fig.~\ref{fig:data_coverage_D16}.
The lines representing the redshift path length are colored based on the sightline's mean SNR per resolution element.
% VK: Check if it was median or mean or something else. Also SNR per pixel or per resolution element? 
%%TH revised
The gaps in the spectra correspond to masked regions. 
It is noticeable that some gaps appear at the same wavelength (i.e same Ly$\alpha$ redshift)
for different sightlines, which is because these are metal absorbers in the \ac{MW} and a geocoronal O~{\sc i} emission line at wavelength $\lambda =1300$ \AA.
% VK it's the largest gap in the spectrum, need to write the actual wavelength
% VK need concluding line here to the subsection
% VK adding below - modify as you wish
After selecting the spectral regions containing \lyaf{} and metal masking, we perform VP-fitting using the automated code as described in the following subsection. 

% VK This is very similar to a figure from my 2019 paper. somewhere we need to mention this  noting that 2019 paper used only 65 QSOs whereas you are using all 82. This is correct right? Please check. 
%%TH yes this is new plot I made, there are slightly more qsos
% VK: Another thing is you need to mention that the dots represent the emission z and for the 5 QSOs on right side the z is beyond 0.5 and hence not in the plot.
%%TH revised in the caption

\subsection{Voigt-Profile Fitting}
\label{sec:vpfit_z01}

In this work, we use our automated line-fitting program based on \vpfit{}\footnote{VPFIT code: \url{http://www.ast.cam.ac.uk/~rfc/vpfit.html}} version 11.1  \citep{vpfit}, 
which fits Voigt profiles convolved with the instrument LSF to \lya{} lines. 
We employ a fully automated Python wrapper adapted from \citet{Hiss2018}, which controls \vpfit{} with the help of its front-end and back-end programs \texttt{RDGEN} and \texttt{AUTOVPIN}, 
to consistently fit both observations and mock spectra from simulations. 
We set up \vpfit{} to explore the range of parameters $1  \leq  b  \leq 300~\text{km/s} $ and $11.5 \leq \log ( N_{\mathHI} / \text{cm}^{-2}) \leq 18$ for every single \lya{} absorption line. 
\vpfit{} automatically varies these parameters and fits for additional component lines until the $\chi^2$ with respect to the whole spectral segment is minimized. 
Such a VP-fitting procedure is applied to the whole spectral segment in observations, 
fitting both the \lya{} lines and metal lines, 
including both intervening metal lines and those from interstellar medium of \ac{MW}; for simplicity, 
hereafter we refer to these collectively as metal lines.
We use the metal masks for the removal of these metal lines as later discussed in \S~\ref{sec:metal_ID_z01}.

During our VP-fitting procedure, we observed weak artificial lines in the HST COS spectra that were absent from our forward-modelled mock spectra. A visual assessment suggests that these minor features are artefacts arising from issues like
% VK is flat fielding really a reason ? 
%%TH I am not sure, so I revised it.
continuum placement, or data reduction, especially in spectra with high \ac{SNR}. 
Therefore, we introduced a fixed 'floor' of 0.02 in quadrature to the normalized flux noise vector across all spectra, effectively adding robustness without increasing the noise in the normalized flux.
%This adjustment was derived 
We arrived at this adjustment
through trial and error, according to the presence of absorption lines with low $b$ and $N_{\mathHI}$ identified by \vpfit{} in the highest \ac{SNR} spectra. These shallow, 
% VK replace faint with shallow 
%TH done
narrow lines, absent in our simulated and forward-modelled sightlines, primarily affect lines with $\log \NHIt{}/\text{cm}^{-2}<12.5$ in our dataset, which are excluded from our inference. For consistency, we applied the same noise floor to the simulated datasets in data processing (see \S~\ref{sec:FM}).

 \begin{figure*}
\includegraphics[width=\textwidth]{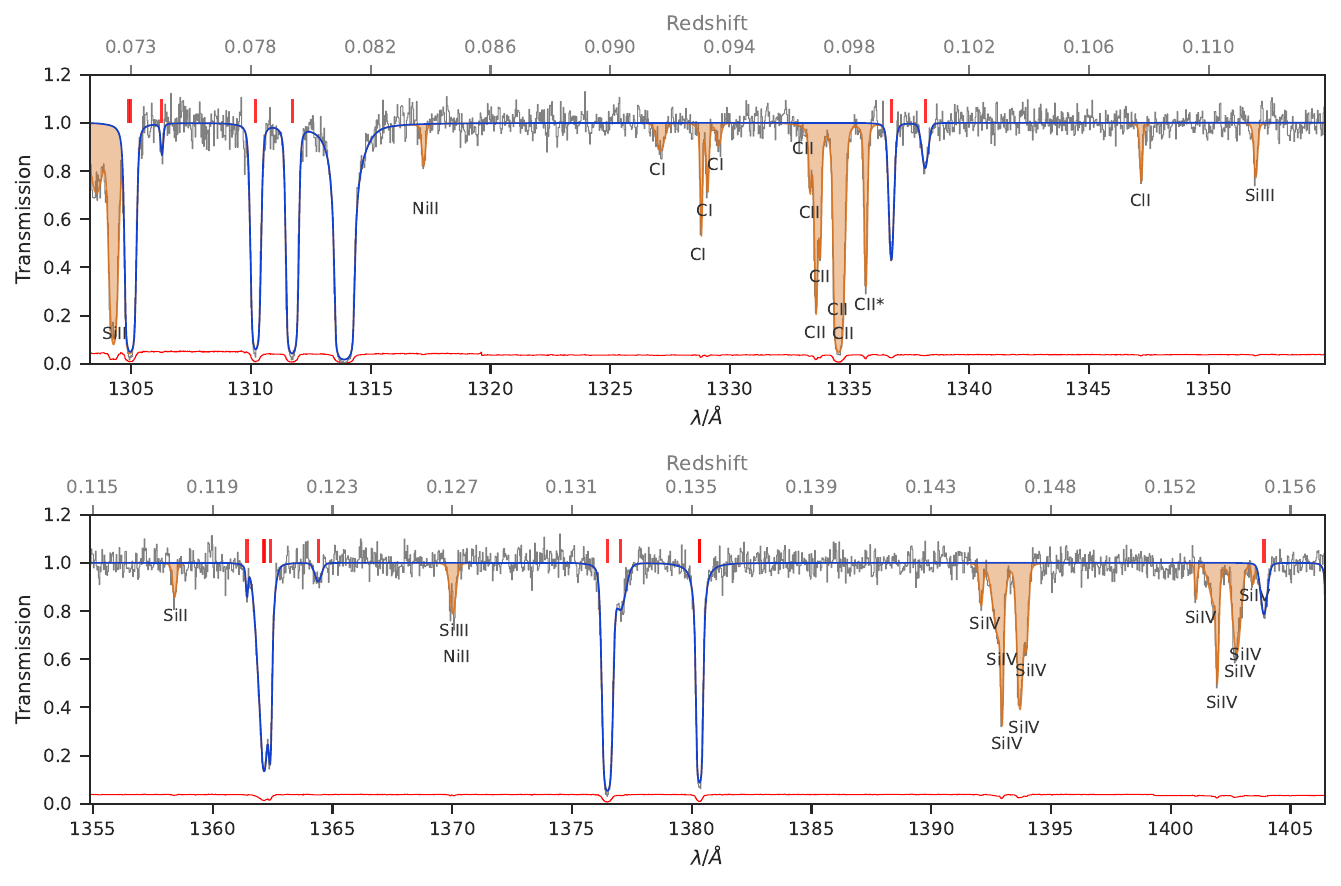}
  \caption{ A segment of HST COS quasar spectrum, sightline of PHL1811, from 1305 to 1405 \AA, with flux shown in grey and noise plotted in red. Orange shaded regions indicate metal line identification and masks, and the fit models of \lya{} forest are shown in blue, while locations of the corresponding \lya{} lines are indicated by red vertical lines.}
  % VK: red is used twice - error and here as well. maybe use green for one (maybe for error) so that it is easy to see
  % VK: the gray color labels on the top axis don't look good. Keep them dark.
  %The fitting procedure is done by the automated program \vpfit, with the corresponding COS LSF taken into account.
  % VK commenting above line -- not needed since text explained it well }
  \label{fig:OB_VP_z01}
\end{figure*} 

%Our \vpfit{} wrapper is designed to incorporate the custom \ac{LSF}, which is important for COS due to its non-Gaussian \ac{LSF}. In addition, the effective resolution of the HST COS gratings also depends on its lifetime position during the observations, which is also taken into account in our VP-fitting program\footnote{Although our \vpfit{} wrapper is compatible with \ac{LSF} in \vpfit{}, only a single \ac{LSF} without wavelength dependence can be used at once. As such, for the input into \vpfit{}, we use the \ac{LSF} at the lifetime position of the data and evaluate it at the central wavelength of the spectrum that we are trying to fit. }, as well as in our forward modelling procedures (see \S\ref{sec:FM}). The VP-fitting program is applied to both observed and simulated spectra so as to make sure our statistics are not biased.

Our \vpfit{} wrapper is specifically designed to incorporate the custom \ac{LSF}, which is critical for COS data analysis because the COS \ac{LSF} is significantly non-Gaussian. In addition, the effective spectral resolution of the HST COS gratings depends on the lifetime position of the instrument at the time of observation. This dependence is explicitly included in our VP-fitting program.\footnote{While our \vpfit{} wrapper is fully compatible with \vpfit{}’s built-in treatment of \ac{LSF}s, only a single, wavelength-independent \ac{LSF} can be applied at once. Therefore, for each input spectrum, we adopt the \ac{LSF} corresponding to the relevant lifetime position and evaluate it at the central wavelength of the spectral region being fitted.}
Both the VP-fitting routine and our forward-modelling framework (see \S\ref{sec:FM}) consistently account for these instrumental effects. To maintain consistency, we apply the same VP-fitting program to both observed and simulated spectra, ensuring that our statistical inferences were not biased by differences in resolution or line-spread function treatment.

The \citetalias{Danforth2016} \lyaf{} quasar spectra consists of sightlines from both COS G130M and G160M grating, which have wavelength coverage 1135 $\sim$ 1450 \AA{} 
% VK: check the lower wavelength - Is it really 900? I remember it was around 1030 
%%TH it is 900 AA on https://hst-docs.stsci.edu/cosihb/chapter-13-cos-reference-material/13-3-gratings/fuv-grating-g130m
%%Danforth use 1135 instead, I revised it to keep the same as Danforth
and 1360 $\sim$ 1775 \AA{} respectively. Some of these spectra are co-added across both gratings, making the actual LSF/resolution ambiguous 
% VK: use alternate to work "complicated". What about ambiguous?
%TH revised
at the overlapping wavelength. After inspecting the flux and noise of individual exposure covering the overlapping region, we decide to VP-fit the spectral segments at $0.06<z<0.16$ using G130M LSF only and fit the spectra in all other bins using the G160M LSF only. In other words, we fit all absorption lines with wavelength below 1410\AA{} with the G130M, and fit lines with wavelength above 1410\AA{} with the G160M. 
In practice, such an arrangement only affects the redshift bin centering at $z=0.2$, and it prevents us from chopping the spectra in the middle of the redshift bin, which causes more edge effects, inducing potential errors in VP-fitting procedure.
% VK: causes more edge effect is not easy to understand. did we discuss edge effect before? If so then keep it as is, if not probably explain the edge effect or revise the last line in the text to make it clear.
%%TH yes I mentioned it one, but I revised it slightly to make it more clear
%In addition, it helps us achieve the best fit for the absorption feature, i.e., obtaining the lowest $\chi^2$ for the fitting, although the difference is minor.
This arrangement is also applied to our forward modelling procedures (see \S\ref{sec:FM}). 

%% JFH can you make it more clear if data from both setups were actually co-added because if so, you don't really know the resolution since it may depend on the S/N weights used in the co-addition. 
%%TH these spectra are co-added. In this case we can't model them 100% correctly. I personally don't know how to address or justify this. It seems to be an error in this work, although the actual effects are limited.

Furthermore, following conventions established in previous studies \citep{schaye2000,rudie2012,Hiss2018}, we apply additional filtering criteria for both $b$ and $N_{\mathHI}$ in this study. We restrict our analysis to $b$-$N_{\mathHI}$ pairs within the range of $12.5 \leq \log (N_{\mathHI} / \text{cm}^{-2}) \leq 14.5$ and $0.5 \leq \log (b / \text{km s}^{-1}) \leq 2.5$. Such a specific range is chosen to ensure the absorption lines are not strongly saturated while ensuring the majority of \bndist{}s from all our Nyx simulation models are included in the $b$ and $N_{\mathHI}$ ranges (see \S~\ref{sec:Thermal_para}).
It also enhances sensitivity to the IGM thermal state and reduces the impacts of poorly understood strong absorption lines, which arise predominantly from the circumgalactic medium (CGM) of intervening galaxies.
% VK: true!

One of our COS spectra, sightline of PHL1811, 
% VK phl1811 is just a shorthand name of the quasar, write it porpery "One of our COS spectra, of a quasar sightline xxxx," Follow the same comment on fig. 2 captions
%%revised
and the corresponding VP-fitting results with metal masking are shown in Fig.~\ref{fig:OB_VP_z01}. 
The continuum normalized spectrum is plotted in gray, and the fit to the unmasked spectrum consisting of the identified \lya{} absorbers is shown in blue. The \lya{} lines used for our \bn{} dataset (after applying all filters) are indicated by red vertical lines. 
%to the unmasked spectrum is shown in blue and consists of the identified \lya{} absorbers. 
The parts of the fitted spectrum shown in orange illustrate the masked segments based on the metal identifications reported in \citet{Danforth2016}. The following subsection outlines our metal identification and masking procedure. 
% VK added a concluding line above

\subsection{Metal Identification}
\label{sec:metal_ID_z01}

As previously mentioned, our VP-fitting procedure fits all absorption lines, including \lya{} and metal lines. For our analysis, which focuses on the \bn{} of the \lyaf{}, it is crucial to exclude these metal lines. We utilize archival metal identification data presented in \citetalias{Danforth2016}. For each spectrum, we create a mask around each identified metal line, which initially centers on the reported wavelengths of the metal lines and has a default width of $\Delta v=50$ km/s, which is chosen based on the resolution of COS, i.e., $15 \sim 20$ km/s for this dataset.
We then refine these masks according to our VP-fit results, which is needed since our VP-fitting results do not all match the  \citetalias{Danforth2016} metal IDs precisely, due to the different VP-fitting procedures used in this work compared to \citetalias{Danforth2016}. 
We start by locating absorption regions where the fitted normalized flux, $F_\text{line,fit}$, is less than or equal to 0.99 (as depicted by the blue line in Fig.~\ref{fig:OB_VP_z01}). If an absorption line region overlaps with a mask, we extend the mask's width to fully cover the line, using the full width at half maximum (FWHM) of the detected line, calculated as $2\sqrt{\ln 2} b$, where $b$ is determined by \vpfit{}. We then manually adjust the masks to bridge small gaps ( $\lesssim$ 50 km/s) between masked regions to ensure full coverage of all lines close to the identified metal lines to ensure that all potential contamination is eliminated. Additionally, we manually mask regions affected by Damped \lya{} absorption systems (DLAs), which disrupt \vpfit{} results. 
% VK: Are there any DLAs? I think there are none. Maybe remove this line then.
These post-processing masks are created after initially applying \vpfit{} under the assumption that all absorption lines are \lya{}, and later, any absorption lines within these regions are excluded from our \bn{} dataset.
The pathlength covered by the metal masks are subtracted from our total pathlength, resulting in a net \lya{} pathlength of $\Delta z=$4.43 for all four redshift bins. %The complete set of quasar segments and their corresponding masks are detailed in Appendix~\ref{sec:data_masks_z01}.
%% JFH Looks like this appendix is not there yet. 
%% TH that would be too long so I reemoved them.

With our imposed filters on the \bn{}, 
we find that 84 out of 741 lines are masked for our whole sample, and that leaves us with a \bn{} dataset consisting of 657 \lya{} absorption lines. 
We divide the 657 \lya{} absorbers into four redshift bins: $0.06 < z <0.16$, $0.16 < z < 0.26$, $0.26 < z < 0.36$, and $0.36 < z <0.48$ respectively, according to their central wavelength as given by \vpfit{}. 
Such ranges for these bins are selected to ensure their pathlength-weighted redshift centres at $z=0.1, 0.2, 0.3,$ and $0.4$,
respectively, matching the redshift of our simulation snapshots.
This provides us with the number of \lya{} lines to be 270, 201, 102 and 84 and redshift pathlength $\Delta z=$ 1.79, 1.30, 0.78, and 0.56 in the bins centred at  $z=0.1$, $0.2$ $0.3$ and $0.4$, respectively.
In Table~\ref{tab_abs_z01} we summarize our \bn{} dataset for each redshift bin, with 
redshift pathlength, number of final \lya{} lines as well as 
median values for the $b$ and $N_{\mathHI}$.
\begin{table}
\caption{Summary of the of the observational dataset}
\centering
\renewcommand{\arraystretch}{2}
\begin{tabular}{ccccc}
\hline
$z$ bins  & $\Delta z$ & Number   & $b_\text{m}/ \text{km s}^{-1} $ & $\log (N_{{\mathHI},\text{m}} /  \text{cm}^{-2})$ \\ 
\hline
$0.06 \leq z \leq 0.16 $ &1.79  & 270 &34.27 & 13.22\\
%\hline
$0.16 < z \leq 0.26 $ &1.30  & 201 &36.05 & 13.14\\
%\hline
$0.26 < z \leq 0.36 $ &0.78  & 102 &32.44 & 13.30\\
%\hline
$0.36 < z \leq 0.48 $ &0.56  & 84 &32.29 & 13.31\\
\hline
\end{tabular}
\raggedright
\begin{tablenotes}
      \small
      \item The numbers of identified \lya{} lines in each redshift, the total pathlength $\Delta z$, 
and the median value $b_\text{m}$ and $\log N_{{\mathHI},\text{m}}$. 
    \end{tablenotes}
\label{tab_abs_z01}
\end{table}
% VK formatted better to remove hline and more space in between
% VK remove the caption below --- replace with notes otherwise, it shows two table numbers

\section{Simulations}
  \label{sec:simulations}
  
 \begin{figure*}
 \centering
\includegraphics[width=1.0\textwidth]{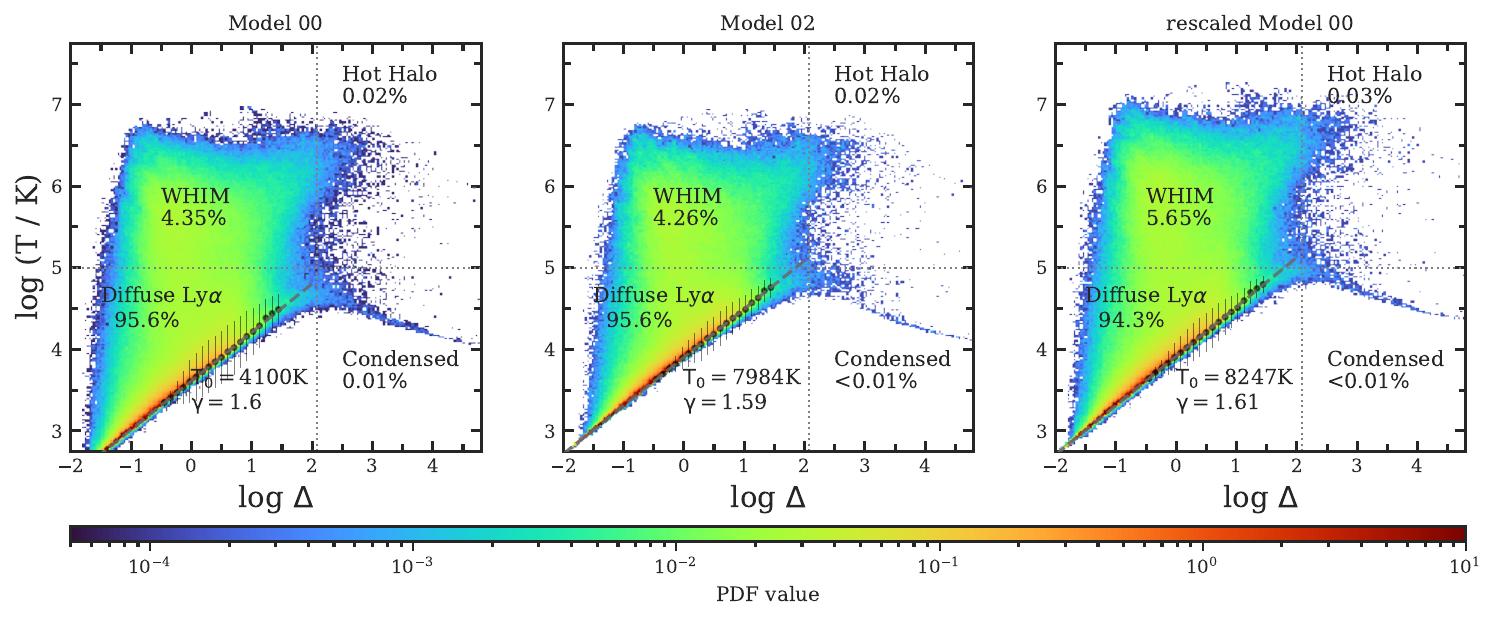}
  \caption{ Volume-weighted $T$-$\Delta$ distribution for four simulation models at $z=0.1$. The left panel presents the Nyx model 00 with $T_0 =4100$ K and $\gamma = 1.60$, while the middle panel features the Nyx model 02 with $T_0 = 7984$ K and $\gamma = 1.59$. The right panel displays a model post-processed by doubling the temperature in model 00, resulting in $T_0 = 8247$ K and $\gamma = 1.61$, as determined by our $\Delta-T$ fitting procedure. The best-fit power-law relationships are depicted as grey dashed lines. The $\log T$ for each bin is represented by black dots, accompanied by 1-$\sigma_T$ error bars in black. Volume-weighted gas phase fractions are annotated. 
  The gas in the simulation is divided into four phases based on the temperature and density ($T=10^5$~K and $\Delta$ = 120), namely the \ac{WHIM}, Diffuse Ly$\alpha$, Hot Halo gas, and Condensed.}
  \label{fig:Nyx_Rho_T_z01}
\end{figure*}

We employ a suite of Nyx cosmological hydrodynamic simulations \citep[see][]{Lukic2015, nyx} to model the low-redshift \ac{IGM}. Nyx is a massively parallel cosmological simulation code specifically developed for simulating the \ac{IGM}. In Nyx, dark matter is treated as self-gravitating Lagrangian particles to capture its evolution, and the baryons are modelled as ideal gas on a uniform Cartesian grid using an Eulerian framework. The dynamics of the gas are computed using a second-order piece-wise parabolic method, which ensures accurate effects from shocks.

Nyx includes the key physical processes essential for modelling the \lyaf{}. It assumes the gas to be of primordial composition with a hydrogen mass fraction of 0.76, helium mass fraction of 0.24, and zero metallicity. Processes such as recombination, collisional ionization, dielectric recombination, and cooling are modelled following the methodologies discussed in \citet{Lukic2015}. Additionally, Nyx simulates inverse Compton cooling against the cosmic microwave background and accounts for the total thermal energy loss due to atomic collisional processes. The standard configuration of Nyx uses a spatially uniform UV background from \citet[][]{H&M2012}.  When generating the \lya{} forest in post-processing (see \S\ref{sec:skewers}), the photoionization rates from the UV background are treated as free parameters, enabling us to determine the IGM's ionization state. Since Nyx simulations focus on the IGM, they do not include galaxy formation and relevant feedback mechanisms, which significantly reduces computational demands, helping the execution of a large ensemble of simulations with varying thermal parameters (as detailed in \S\ref{sec:Thermal_para}).

In this work, each Nyx simulation has been initiated at $z=159$ and ran until $z=0.03$. The simulation has a box size of $L_{\text{box}} = 20~{\rm cMpc}\slash h$ with $1024^3$ Eulerian cells for baryons and the same number of dark matter particles. This chosen box size balances the need to manage computational resources while ensuring that the results converge to within a margin of $< 10 \%$ on small scales. A more detailed discussion on the considerations regarding resolution and box size is provided in \citet{Lukic2015}.

 \subsection{Thermal Parameters and Simulation Grid}
   \label{sec:Thermal_para}

\subsubsection{The THERMAL suite}
\label{sec:THERMAL}

To model various thermal histories of the \ac{IGM}, we use a revised subset of the \ac{THERMAL}\footnote{Details of the \ac{THERMAL} suite are available at http://thermal.joseonorbe.com.} suite of Nyx simulations \citep[also see][]{Hiss2018,WaltherM2019}. From this suite, we select 51 models, each representing a distinct thermal history. For each model, we generate four simulation snapshots at redshifts $z= 0.1, 0.2, 0.3$, and $0.4$ to measure the thermal state, characterized by parameters [$\log T_0$,$\gamma$]. We adjust the photoheating rates ($\epsilon$) to achieve varied thermal histories, following the approach described in \cite{Becker2011}. In this method, $\epsilon$ is modeled as a function of overdensity, represented by the equation:
\begin{equation}
\epsilon = A \epsilon_{\rm HM12} \Delta^B,
\label{eqn:heating}
\end{equation}
where $\epsilon_{\rm HM12}$ is the baseline photoheating rate per H~{\sc ii} ion, as tabulated in \cite{H&M2012}, and $A$ and $B$ are free parameters that allow us to create models with different thermal histories.

As the Universe 
%% JFH I usually make universe capital everywhere. But adopt a convention adn be consistent. 
% VK: yes, mnras insists Universe and not universe 
%TH revised 
evolves towards lower redshifts, the thermal state of the IGM tends to converge to a small range of lower temperatures due to adiabatic cooling dominated by Hubble expansion, which makes it a challenge to generate models with uniformly distributed $T_0$ and $\gamma$ values \citep[See][]{WaltherM2019}. Specifically, creating models with a low $T_0$ (below $10^{3.5}~{\rm K}$) combined with a high $\gamma$ value (above 1.9) at lower redshifts is particularly challenging. Decreasing the photoheating rates to lower $T_0$ results in Hubble expansion cooling becoming dominant compared to photoheating, which typically drives $\gamma$ towards a value close to 1.6, as discussed in \citep{McQuinn2016}. 
% VK: also mention the difficulty in getting higher T_0 -- the photoheating becomes ineffective because of a very low neutral fraction of the gas and therefore it is very difficult to heat the gas to 10^4 K and beyond  --- as explained in the next section in details
%TH revised
Consequently, the $T_0$-$\gamma$ grid has an irregular shape with no models in the high $\gamma$ low $T_0$ regions (see the $T_0$-$\gamma$ grid in Fig.~\ref{fig:corner_z01}). 
This irregularity also arises from the design of the parameter grid in the \ac{THERMAL} suite, which is designed to analyse the thermal state at higher redshifts \citep{WaltherM2019}. 
% VK: motivate the next subsection i.e why we need rescaling models
%TH revised

%% JFH It is not good style to start a section title with a symbol. Instead Just write it out. Rescaled Temperature Models or Resdcaled T0 models. 
%%TH revised
\subsubsection{Models with Rescaled $T_0$}
\label{sec:rescale}

As will be discussed later in \S~\ref{sec:result_z01}, 
our data favour models with high $T_0$ at $z=0.1,0.2,0.3$ and $0.4$. These extremely hot models are challenging to generate at low-$z$ solely by varying the H~{\sc i} photoheating rate (see Eq.~\ref{eqn:heating}), 
%% JFH Reference equation 2 here please. 
%%TH revised
since the IGM is dominated by the adiabatic cooling caused by Hubble expansion at this epoch, and the heat injection caused by H~{\sc i} photoionization fades away quickly  
\citep[see e.g.,][]{Sanderbeck2016} 
% VK: dont'e cite the review paper. The paper with Sanderback has shown the evolution of T0 to low-z
%%TH revised
mainly because of a very small H~{\sc i} fraction. 
As a result, the $T_0$ of the IGM at low-$z$ is insensitive to the H~{\sc i} photoheating.
To address this, we rescale the IGM temperature in our simulations in post-processing to model the IGM with high temperatures, as described below. 

For each redshift bin, we divide the models into $\gamma$ bins with $\Delta \gamma =0.1$ and
select the simulation with highest $T_0$ in each $\gamma$ bins, and multiply their temperature $T$ (at each simulation cell) by a factor $k_{re} = [\sqrt{2},2,2\sqrt{2},4,4\sqrt{2},8]$
%% JFH how were these numbers chosen. 
%%TH just random
respectively to generate $6 \times 13 = 78 $ new models (13 different $\Gamma_{\mathHI{}}$ values). 
% VK: how does it generate 66 new models? Mention it clearly
%%TH revised
The other properties of the simulation remain unchanged, and since we rescaled the temperature of all simulation cells uniformly the whole $\Delta$-$T$ distribution of the simulation model still follows the power law  $\Delta$-$T$ relationship (see Eq.~\ref{eqn:rho_T}) with the $T_0$ rescaled.
The rescaling procedure is depicted in Fig.~\ref{fig:Nyx_Rho_T_z01}, and the distribution of [$T_0$, $\gamma$] for the 
%origianl 
original %% VK2 change
models and the rescaled models is shown in Fig.~\ref{fig:corner_z01}, where the original models are shown as blue dots, and the models with rescaled $T'_0 = k_{re}*T_0 $ are shown in orange. This temperature rescaling procedure is applied to all four redshift bins.

\subsubsection{Measuring the IGM thermal state $[T_0, \gamma]$}
\label{sec:measure_rho_T}

To measure the thermal state for each model, 
we fit a power-law temperature-density ($T$-$\Delta$) relation (see Eq.~\ref{eqn:rho_T}) to the
temperatures and densities in the simulation domain.
While fitting the $T$-$\Delta$ relationship, 
we noticed broader distributions of the \ac{IGM} temperatures in low redshift ($z \lesssim 1.0$) compared to high redshift ($z>3$).
To accommodate the dispersion in the IGM $T$-$\Delta$ distribution while fitting the power-law relationship, we adopt the fitting approach detailed in \citet{Hu2022}.

This method first segregates the diffuse \lya{} gas \citep[ $\log (T/\text{K}) <5$ and $\Delta <120$, see][]{Dave2010} into 20 bins based on $\log \Delta$. 
% VK provide the log delta value
%%TH revised
A linear least squares fit is then applied to the average temperatures within each bin. For this study, we have 
% VK we have, usually short forms of these kinds are not used in the scientific writing 
%%TH revised
adjusted the fitting range to $-0.5 < \log \Delta < 1.5$. 
Examples of the $\Delta$–$T$ distribution, together with the corresponding power-law fits, are presented in Fig.~\ref{fig:Nyx_Rho_T_z01}. In each panel, the best-fit power-law relation is indicated by grey dashed lines. The binned values of $\log T$ are shown as black points, with error bars representing the 1$\sigma_{T}$ uncertainty.  The latter is defined as half of the temperature interval that encloses the central $\pm$ 1-$\sigma$(16\%–84\%) of the probability density within each bin.
%% JFH You need to explain how these "error bars" are defined. You did not do that. 
%%TH revised
The left panel shows the Nyx model denoted as 'model 00'
% VK Nyx model denoted as 'model 00' having  
%%TH revised
with $T_0=4100$ K, $\gamma=1.60$, 
and the middle panel shows the Nyx model denoted as 'model 02'
% VK model denoted as 'model 02' having
%%TH revised
with $T_0=7984$ K, $\gamma=1.59$ generated by varying the parameters $A$ and $B$ in Eq.~\ref{eqn:heating}.
The right panel shows the rescaled model 00 generated by multiplying the temperature in model 00 by two.
It exhibits a $T_0 = 8247$ K and $\gamma =$1.61 according to our $\Delta-T$ fitting procedure.

\subsubsection{ Varying the UVB photoionization rate $\Gamma_{\mathHI{}}$}
\label{sec:Gamma_grid}
To constrain the ionization state of the IGM, we treat the \HI{} photoionization rate, $\Gamma_{\mathHI{}}$, as a free parameter while generating \lyaf{} skewers from our simulations.
As such, we add an additional dimension $\log \Gamma_{\mathHI{}}$ to our parameter grid used for our inference framework,
extending it to [$\log T_0$, $\gamma$, $\log \Gamma_{\mathHI{}}$].
This procedure is carried out during the post-processing phase of the simulation when the simulated sightlines are generated (see \S~\ref{sec:skewers}). In this study, the values of $\Gamma_{\mathHI{}}$ range from $\log (\Gamma_{\mathHI{}} /\text{s}^{-1}) = -13.834$ to $-12.931$, in logarithmic increments of $0.075$ dex, providing a total of 13 distinct values. These values are consistently applied across all redshift bins.

\subsection{Skewers}
\label{sec:skewers}

In this work, we generate mock \lya{} spectra by calculating the \lya{} optical depth ($\tau$) along mock lines-of-sight in the simulation, referred to as skewers for simplicity. For each simulation model, a set of 15000 random skewers is generated aligned with the $x$, $y$, and $z$ axes of the simulation box, distributing 5000 skewers per axis. Properties essential for computing the optical depth are extracted from each cell along these skewers, including temperature ($T$), overdensity ($\Delta$), and the line-of-sight velocity ($v_z$). The hydrogen neutral fraction ($x_{\mathHI{}}$), crucial for synthesizing the \lyaf{}, is determined by assuming ionization equilibrium, which considers both collisional ionization, influenced by the gas temperature, and photoionization.
As presented in \S\ref{sec:Gamma_grid}, $\Gamma_{\mathHI{}}$ is treated as a free parameter during post-processing. Given that Nyx does not perform radiative transfer, we approximately treat the impact of self-shielding of the UV background in optically thick gas. Following the methodology of \cite{Rahmati2013}. This approach involves attenuating $\Gamma_{\mathHI{}}$ in cells with dense gas to simulate the effects of self-shielding.

Based on the extracted values of $x_{\mathHI{}}$, $T$, $\Delta$, $v_z$, and $\Gamma_{\mathHI{}}$, we calculate the optical depth $\tau$ in redshift space by summing the contributions of all cells along the line-of-sight in real space, employing the full Voigt profile approximation as outlined by \citet{Tepper-G2006}. The continuum normalized flux of the \lyaf{} along these skewers is then calculated using $F = e^{-\tau}$.
This procedure is repeated for each specified value of $\Gamma_{\mathHI{}}$ to generate skewers. %corresponding to the given $\Gamma_{\mathHI{}}$
% VK commented the redundant part
%
%% VK2 change 
%In this work, we do not adopt the common practice of rescaling $\tau$ for different $\Gamma_{\mathHI{}}$ values, a method typically employed at higher redshifts. This is due to the unique characteristics of the low-$z$ IGM. Contrary to the high-$z$ IGM, which is predominantly influenced by photoionization, the low-$z$ IGM includes a significant proportion of shock-heated WHIM. %Warm-Hot Intergalactic Medium (WHIM) gas. 
%
%Therefore, it is essential to recalculate the skewers to account for the contributions from collisionally ionized gas (See \citet{Khaire2019} for detailed discussion).
In this work, we do not adopt the common practice of rescaling $\tau$
for different $\Gamma_{\mathHI{}}$ values, a method typically employed at
higher redshifts. As demonstrated in \citet[][their figure~4]{Khaire2019},
this rescaling biases the recovered Ly$\alpha$ forest flux power spectrum,
with a scale-dependent deviation that grows as the difference between the
initial and target $\Gamma_{\mathHI{}}$ increases. The column density
distribution is considerably less sensitive to this approximation,
consistent with the test presented in \citet[][their figure~B3]{Bolton2022};
that test, however, spans only a limited range in $\Gamma_{\mathHI{}}$.
In our analysis the likelihood makes use of the line abundance
$\mathrm{d}N/\mathrm{d}z$ in addition to the $b$--$N_{\mathHI{}}$
distribution, and our inference grid spans a wide range of
$\Gamma_{\mathHI{}}$ (a factor of $\sim 8$).
We therefore recompute the skewers self-consistently for each
$\Gamma_{\mathHI{}}$ value rather than rescaling $\tau$, which correctly
captures the response of the ionization state, including the
collisionally ionized and self-shielded gas, across the full grid.

\subsection{Forward Modeling of Noise and Resolution}
\label{sec:FM}
 
In this paper, we attempt to constrain the thermal and ionization state of the IGM at $z < 0.5$ based on HST COS data.
Theoretically, the IGM temperatures at mean density $T_0$ is $\sim 5000~{\rm K}$ at $z \sim 0$,
the resulting $b$-values for pure thermal broadening (i.e. the narrowest lines in the Ly$\alpha$ forest) are $b \sim 9$~{\rm km/s}, corresponding to a full width at half maximum (FWHM)$ \sim 15$ km/s.
Such absorption features can not be fully resolved by \ac{COS}, which has a resolution of $15 \sim 20$ {\rm km/s} ($R\sim 15000-20000$), depending on the wavelength and grating.
%% JFH Can you quote the R-values as well please. 
%%TH done
Therefore, it is crucial to carefully model instrumental effects. We forward model the noise and resolution to ensure that our simulation results are statistically comparable with the observational data.

We generate mock datasets with properties consistent with our \citetalias{Danforth2016} \lya{} quasar spectra, which comprise both COS G130M ($R\sim 15000$) and G160M ($R\sim 20000$) quasar spectra. 
%% JFH remind the reader again of the typical resolution of each grating in parenthessis quoted as both an R-value and a FWHM please. 
%%TH revised
As mentioned in \S\ref{sec:vpfit_z01}, for simulated spectra at $0.06<z<0.16$, we forward-model them using the G130M grating ($1135\sim1450$ \AA{}), and for the other three 
bins, we forward-model the mock spectra using the G160M grating ($1360\sim1775$ \AA{}). 
These LSFs are obtained from the Python package {\tt linetools}\footnote{For more information, visit \url{https://linetools.readthedocs.io}.}, 
%% JFH provide a hyperlink to the github repository as a footnote. 
%%TH revised
which also takes into account the HST life position (LP).

For a selected observational quasar spectrum from our dataset, we begin by stitching randomly selected skewers together without repetition to match the initial and final wavelength of the selected spectrum segment. 
%% JFH This is incorrect. In the forward modeling, you should always be resolution convolving at the native simulation resolution and then rebinning onto the observed wavelength grid. You should not be rebinning first, since then you perform the convolution with an unnnecessarily corase pixel scale. Perahps it is not a big effect, but is this really how you do the forward modeling?
% VK: I think the forward model code convolved first and then rebins. @Teng please check
%%TH yes it convolved first and then rebins. I have revised the text
Following this, we then convolve the simulated spectra with the HST \ac{COS} LSF, taking into account the specific grating and LP used for each observed spectrum. The \ac{COS} \ac{LSF} is tabulated for 50 pixels in each direction, and we interpolate this \ac{LSF} onto the wavelengths of the mock spectrum to achieve a wavelength-dependent \ac{LSF}. Each output pixel is modelled as a convolution between the stitched skewers and the interpolated \ac{LSF} at that particular wavelength.
Afterward, the newly generated spectrum is interpolated to match the wavelength grid of the selected \ac{COS} spectra. 
%% JFH This description is unclear. You state that you rebin above, before convolution, now you state that you convolve then interpolate. I'm confused. This needs to be clarified. 
%%TH it convolved first and then rebins. I have revised the text
The noise vector from the observed quasar spectrum is propagated to our simulated spectrum on a pixel-by-pixel basis by sampling from a Gaussian distribution with a standard deviation equal to $\psi_i$, where $\psi_i$ represents the noise vector value at the $i^\text{th}$ pixel. To prevent any artificial effects on our VP fits during post-processing, as discussed in \S\ref{sec:vpfit_z01}, a fixed noise floor of 0.02 is added in quadrature to the error vector for all simulated spectra.

\begin{figure*}
\includegraphics[width=\textwidth]{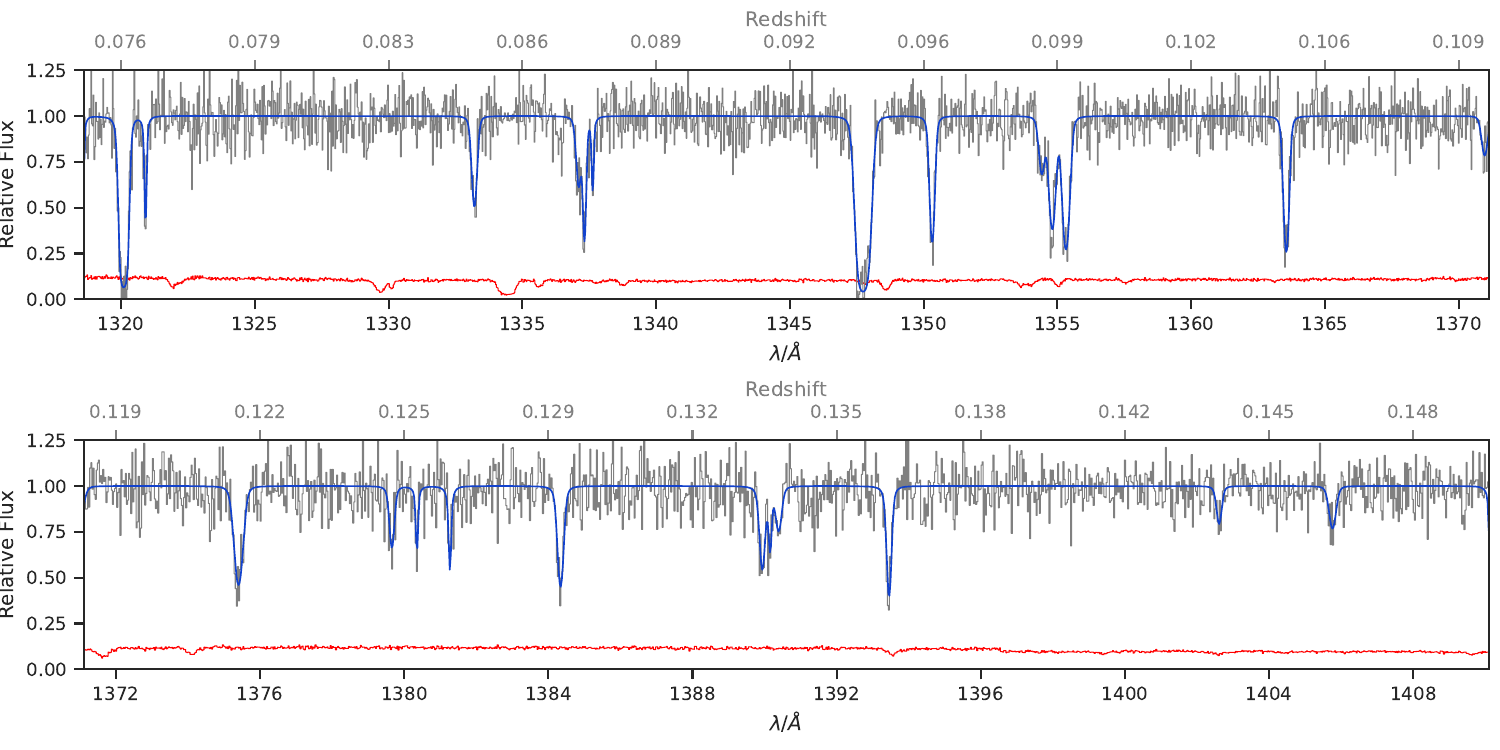}
  \caption{ A segment of a mock spectrum forward modelled based on one of the HST COS spectra. The flux is shown in grey and the noise is plotted in red. The fit models of \lya{} forest are shown in blue. The fitting procedure is done by the automated program \vpfit, with the corresponding COS LSF taken into account.}
  \label{fig:model_VP_z01}
\end{figure*}

For each model, including both Nyx models from the THERMAL suite and those generated by rescaling the temperature, we produced 1000 mock spectra from the 15000 available raw skewers\footnote{
To generate 1000 spectra, approximately 10000 raw skewers are randomly selected from the total pool of 15000 skewers for each model.}. The total pathlength covered by the dataset for each model is approximately $\Delta z_\text{tot} \sim 60$. We then fit Voigt profiles to each absorption line in the spectra to compile the \bn{} dataset, which is used for training the \bndist{} emulator, as will be discussed in \S~\ref{sec:emu}.
For illustrative purposes, an example of a forward-modeled mock spectrum is shown in Fig.~\ref{fig:model_VP_z01}, where the simulated spectrum is shown in grey, the fitted model spectrum from \vpfit{} is shown in blue, and the noise vector is depicted in red.

 \section{Inference Method}
  \label{sec:inference}

 \subsection{Emulating the \bn{} Distribution}
 \label{sec:emu}
 
In this work, we make use of the inference framework following \citet{Hu2022}, which measures the thermal and ionization state [$T_0$,$\gamma$,$\Gamma_{\mathHI{}}$]
% VK Gamma_HI already defined
%%TH revised
of the low redshift \ac{IGM} using its \bndist{} and absorber line density \dndz{}.
The \bndist{} emulator is built on \ac{DELFI}, which turns inference into a density estimation task by learning the distribution of a dataset as a function of the labels or parameters \citep{papamakarios2016, Alsing2018, papamakarios2018, Lueckmann2018, Alsing2019}.
Following \citet{Hu2022}, we make use of \texttt{pydelfi}, the publicly available \texttt{python} implementation of \ac{DELFI},\footnote{See https://github.com/justinalsing/pydelfi} which makes use of \ac{NDE} to learn the  
conditional probability distribution $P(\mathbf{d} \given \boldsymbol{\theta})$ of the data summaries $\mathbf{d}$, as a function of labels/parameters $\boldsymbol\theta$, from a training set of simulated data.
Here the data summaries $\mathbf{d}$ are [$ \text{\NHI{}}$, $ b$], 
and our set of label parameters $\boldsymbol{\theta}$ are the IGM thermal and ionization state [$ T_0$, $\gamma$, $ \Gamma_{\mathHI{}}$].

We generate training datasets by labelling the \bn{} pairs obtained from our 
mock spectra with the aforementioned labels.
We then train the neural network on the summary-dataset pairs pairs $\{T_0,\gamma,  \Gamma_{\mathHI{}} \} - \{b, \text{\NHI{}}\}$.
%% JFH2 This last bit of notation is weird, makes it seems like you lael individual lines, not datasets. Rather than summary-parameter pairs, I would say summary-dataset pairs. I would also use {T_0, gamma, \Gamma}-{b,NHI} which is better notation. 
%%TH revised
Our \bndist{} emulator learns the conditional probability distribution $P ( b \mathbin{,} N_{\mathHI{}} \, \given \, T_0, \gamma, \Gamma_{\mathHI{}})$. These conditional \bndist{}s are then used in our inference algorithm, where we try to find the best-fit model given the observational/mock dataset, which is described in the following section. Note that we train our \bndist{} emulator for each redshift bin separately based on the corresponding training datasets.

\subsubsection{Likelihood function}
  \label{3sec:log-likelihood}

In Bayesian inference, a likelihood $\mathcal{L}= P(\mathrm{data}|\mathrm{model})$ is used to describe the probability of observing the data for any given model.
We adopt the likelihood 
% VK write explicitly $\mathcal{L}=  P(b_i,  \text{N}_{\mathHI{},i} \given \boldsymbol{\theta})$ somewhere in thes section
formalism introduced in \citet{Hu2022},
which is summarized as follows,
\begin{equation}
\ln \mathcal{L} = \sum_{i=1}^{n} \ln (\mu_i) - \left(\frac{\text{d} N}{\text{d} z}\right)_{\rm model}\Delta z _{\rm data}, 
\label{eq:likelihood}
\end{equation}
where $\mu_i$ is the Poisson rate of an absorber occupying a cell in the $b$-$N_{\rm HI}$ plane with area $\Delta { \text{N}_{\mathHI{},i}}\times \Delta b_i$, i.e.
\begin{equation}
\mu_{i}=\left(\frac{\text{d} N}{\text{d} z}\right)_{\rm model}\,P(b_i, N_{\mathHI{},i} \given \boldsymbol{\theta})\,\Delta { N_{\mathHI{}}}\, \Delta b\, \Delta z _{\rm data}. 
\label{eq:mu}
\end{equation}
The $P(b_i,  \text{N}_{\mathHI{},i} \given \boldsymbol{\theta})$ in the equation
is the probability distribution function at the point $(b_i, N_{\mathHI{},i})$
for any given model parameters $\boldsymbol{\theta}$
evaluated from the DELFI \bndist{} emulator.
The $\Delta z_{\rm data}$ is the total redshift pathlength covered by the quasar
spectra from which we obtain our \bn{} dataset, and $\left({\text{d} N}\slash{\text{d} z}\right)_{\rm model}$ is the absorber density which is evaluated for any given set of parameters using a Gaussian process emulator (based on \texttt{George}, see \citealt{Ambikasaran2016}), which is also trained on our training datasets obtained from the Nyx simulation suite. 
More information on the likelihood function and the DELFI emulator, and gaussian process dNdz emulator
%% JFH on the likelihood function, DELFI emulator, and gaussian process dNdz emulator can be found in ...
%%TH revised
can be found in \citet{Hu2022}.

\section{Results}
\label{sec:result_z01}
\begin{figure*}
 \centering
    \includegraphics[width=.80\textwidth]{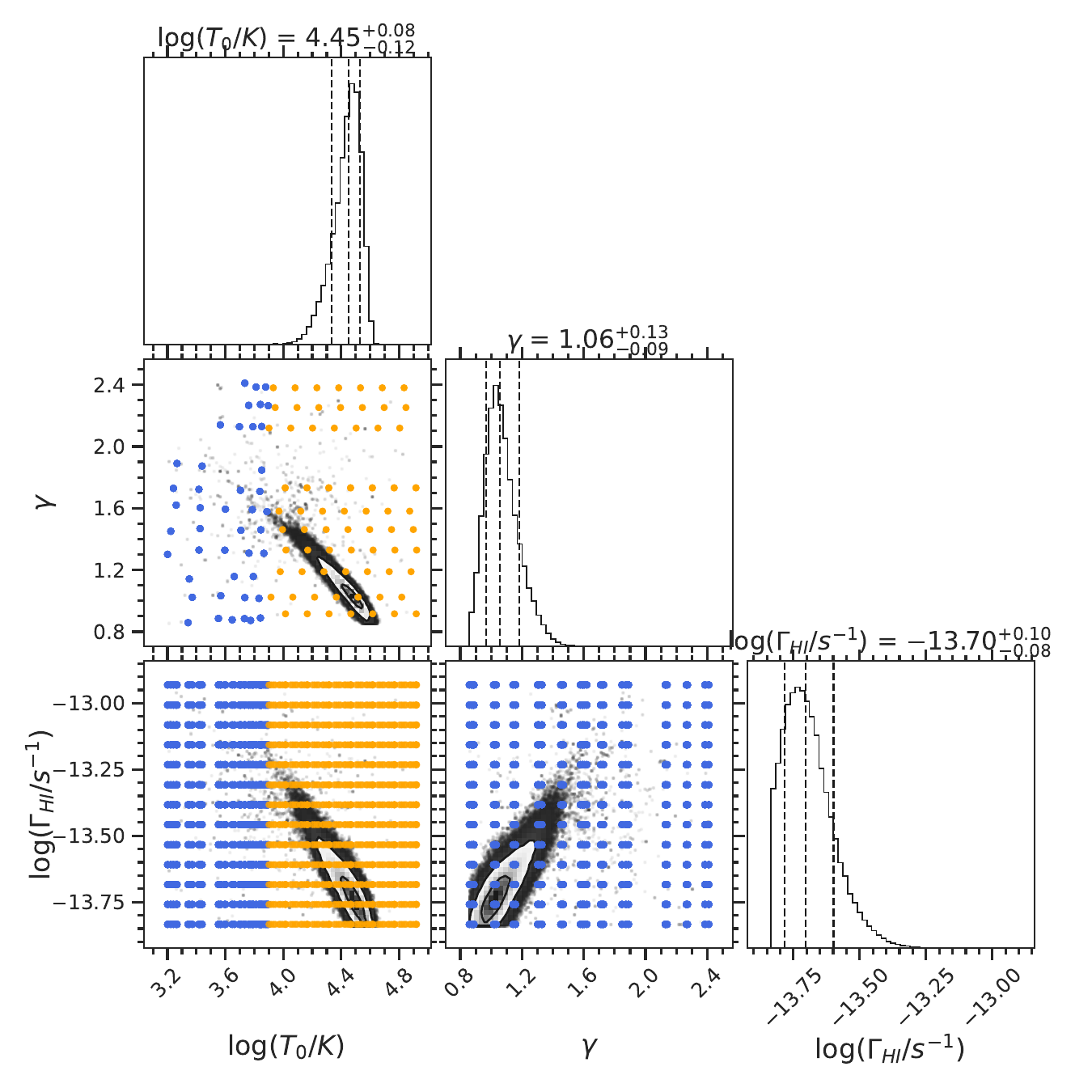}
  \caption{ The MCMC posterior obtained by our inference method using our \bn{} dataset at $z=0.1$. Projections of the thermal grid used for generating models are shown as blue dots.
  The THERMAL Nyx models are plotted as blue dots, and the models with rescaled temperature are shown as orange dots.
  The inner (outer) black contour represents the projected 2D 1(2)-sigma interval. The dashed black lines indicate the 16, 50, and 84 percentile values of the marginalized 1D posterior. }
  \label{fig:corner_z01}
\end{figure*}

\begin{figure*}
 \centering
    \includegraphics[width=.80\textwidth]{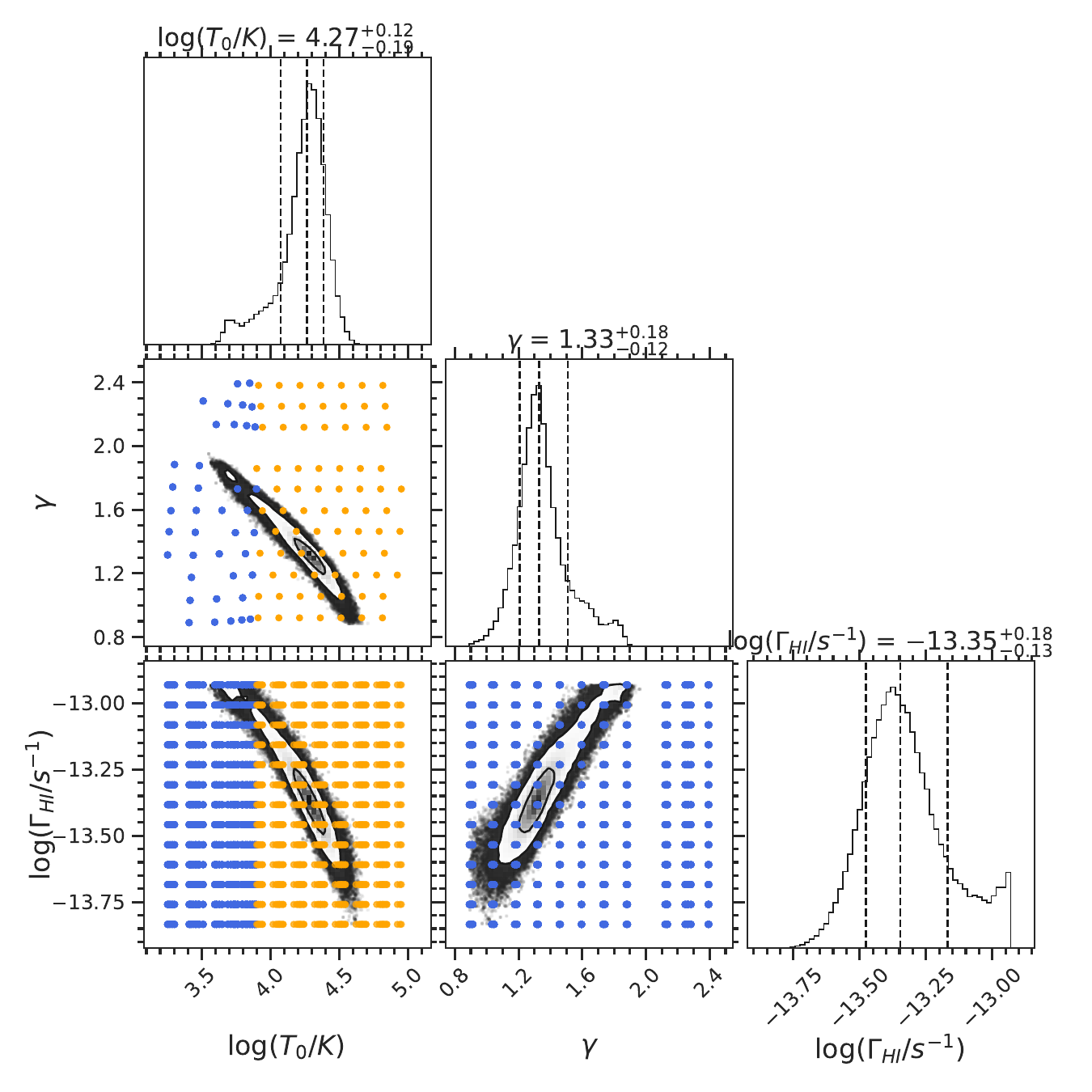}
  \caption{ The MCMC posterior obtained by our inference method using our \bn{} dataset at $z=0.2$.  See Fig. \ref{fig:corner_z01} caption for details.}
  \label{fig:corner_z02}
\end{figure*}

\begin{figure*}
 \centering
    \includegraphics[width=.80\textwidth]{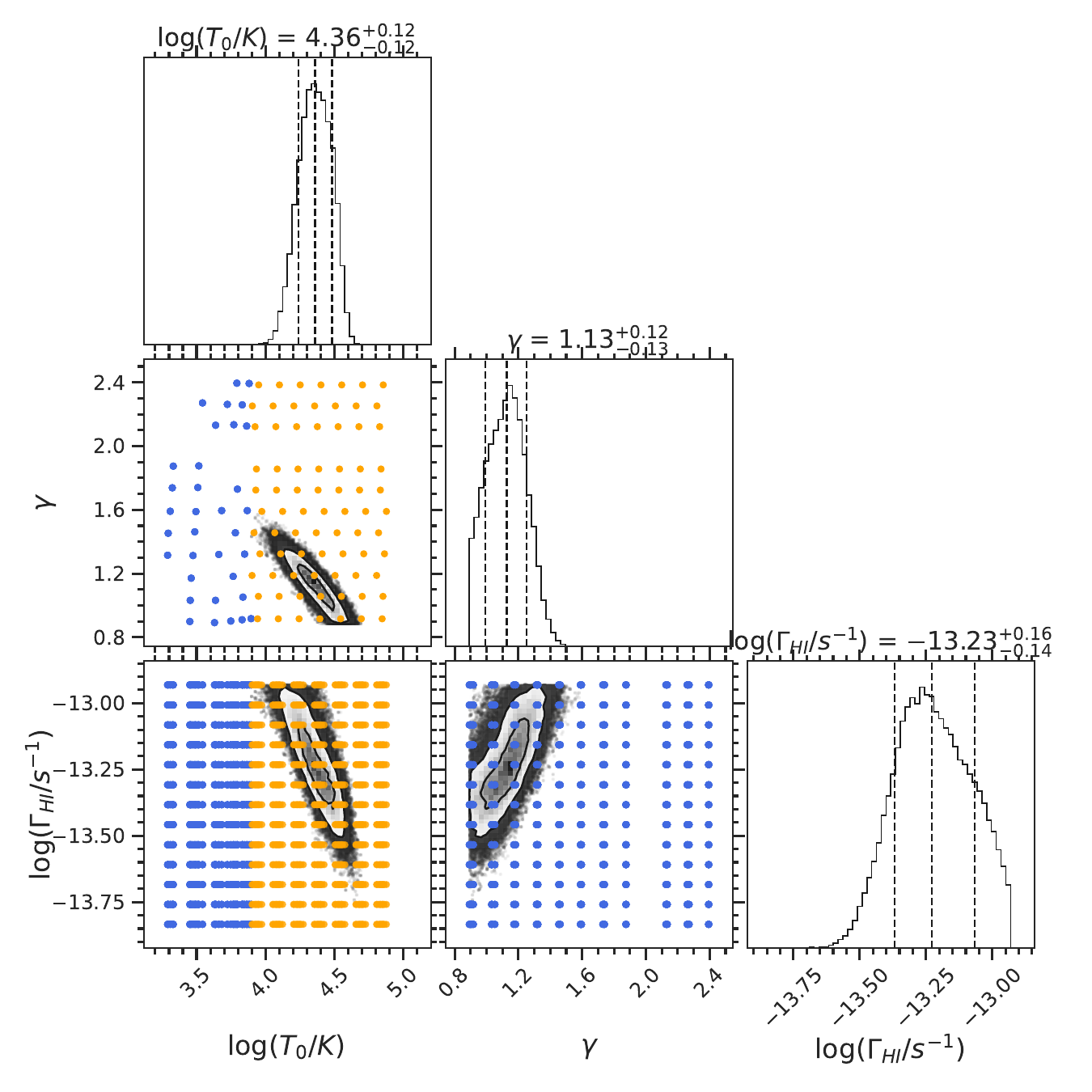}
  \caption{ The MCMC posterior obtained by our inference method using our \bn{} dataset at $z=0.3$. See Fig. \ref{fig:corner_z01} caption for details. }
  \label{fig:corner_z03}
\end{figure*}

\begin{figure*}
 \centering
    \includegraphics[width=.80\textwidth]{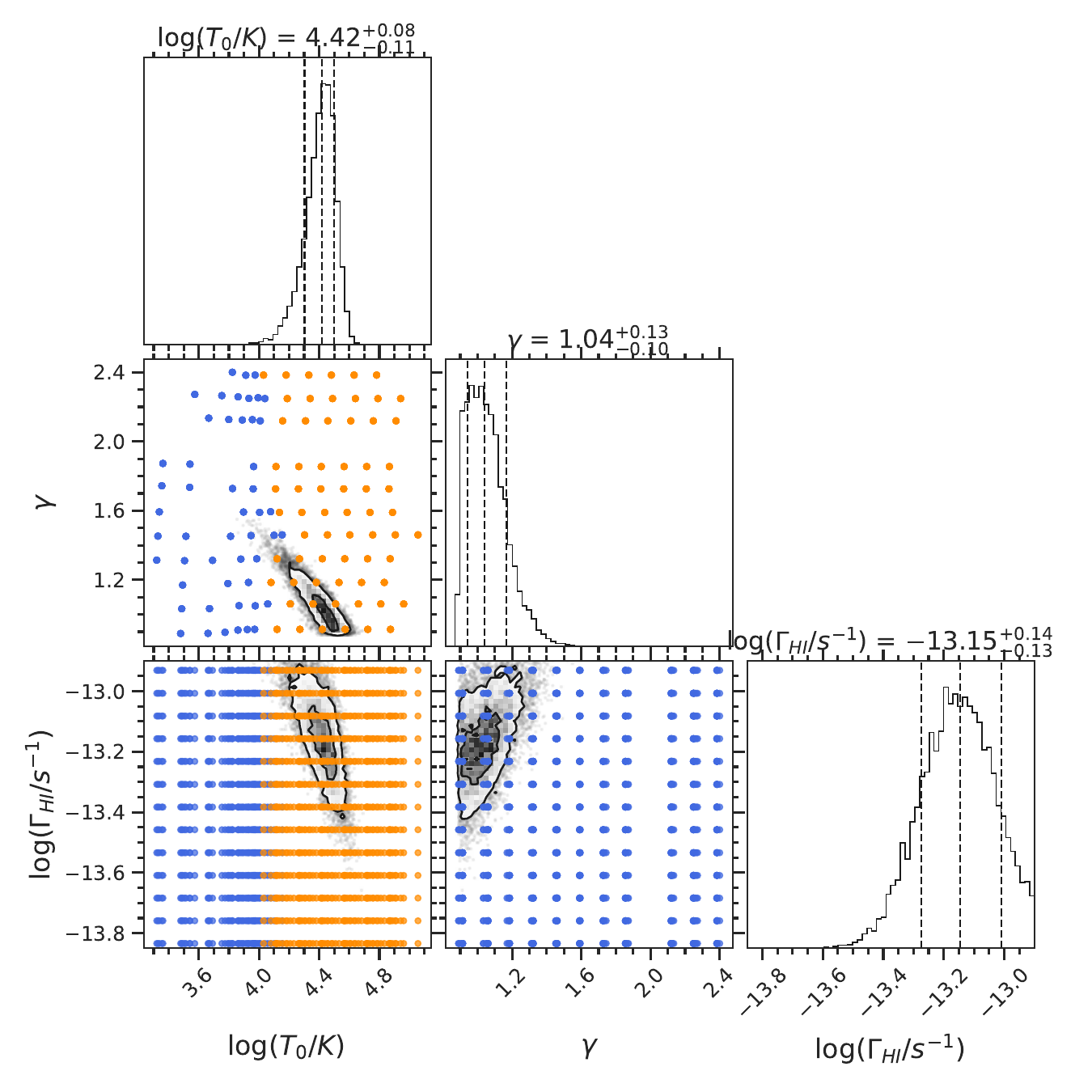}
  \caption{ The MCMC posterior obtained by our inference method using our \bn{} dataset at $z=0.4$. See Fig. \ref{fig:corner_z01} caption for details.  }
  \label{fig:corner_z04}
\end{figure*}

We applied the aforementioned inference method to our dataset in four redshift bins to measure the IGM thermal and ionization state at $z=$ 0.1, 0.2, 0.3 and 0.4. The resulting MCMC posteriors are presented in Fig.~\ref{fig:corner_z01}-\ref{fig:corner_z04} respectively.
%% JFH Just say Fig 5-8 respectively. This reads awkwardly. 
%%TH revised

%% JFH Projections is confusing. The appropriate mathematical term is "marginal". Just stat that the locations of the Nyx models in the parameter space are indicated by ...
%%TH revised
The locations of the Nyx models in the parameter space are indicated by blue dots, and the models with rescaled temperatures are shown as orange dots (see \S~\ref{sec:rescale}). 
The inner (outer) black contour represents the projected 2D 
1(2)-sigma interval. For the histograms, the dashed black lines indicate the 16, 50, and 84 percentile values of the marginalized 1D posterior. 
%% JFH Make it clear you refer to the marginalized 1D distributions shown as histograms for this sentence above.
%%TH revised
Based on the marginalized 2D posteriors, we observe that our results across all redshift bins exhibit the anticipated degeneracies between parameters. Specifically, $T_0$
is degenerate with both $\gamma$ and $\Gamma_{\mathHI}$, which is discussed in \citet{Hu2022}.

%%TH make new plots for this one

We tabulate our inference results (16th, 50th, and 84th percentiles of the marginalized 1D posteriors for each parameter) in
%% JFH You are not just quoting medians but also 16 and 84 precentiles so state that. 
%%TH revised
Table~\ref{tab_inf_result_z01}. 
\begin{table}
\centering
\caption{ Summary of the inference results}
\renewcommand{\arraystretch}{2}
\begin{tabular}{cccc}
\hline
$z$ bins &  $\log(T_0/\text{K})$  & $\gamma$ & log ($\Gamma_{\mathHI} / {\rm s}^{-1}$) \\ 
\hline
$0.06 < z \leq 0.16 $ & ${4.45}^{+0.08}_{-0.12}$ &${1.06}^{+0.13}_{-0.09}$ &${-13.70}^{+0.10}_{-0.08}$\\
%\hline
$0.16 < z \leq 0.26 $ & ${4.27}^{+0.12}_{-0.19}$ &${1.33}^{+0.18}_{-0.12}$ &${-13.35}^{+0.18}_{-0.13}$ \\
%\hline
$0.26 \leq z \leq 0.36 $ & ${4.36}^{+0.12}_{-0.12}$ &${1.13}^{+0.12}_{-0.13}$ &${-13.23}^{+0.16}_{-0.14}$ \\
%\hline
$0.36 \leq z \leq 0.48 $ & ${4.42}^{+0.08}_{-0.11}$ &${1.04}^{+0.13}_{-0.10}$ &${-13.15}^{+0.14}_{-0.13}$ \\
\hline
\end{tabular}
\begin{tablenotes}
      \small
      \item Notes: The inference results i.e., median values of the marginalized 1D posteriors for each parameter, for all four redshift bins. The errors are given by the 1-$\sigma$ error (16-84\%) of the marginalized 1D posteriors.
    \end{tablenotes}
\label{tab_inf_result_z01}
\end{table}
Our inference results show that the temperature of the IGM is much higher than expected and is nearly isothermal, with $T_0$ approaching $30000$K, and $\gamma$ approaching 1.0 at $z=0.1$.  In addition, the $\Gamma_{\mathHI}$ values we inferred are lower than the theoretical model
from \citet{Khaire_Srianand2019}.
%% JFH which result are you referring to? Make it more clear, i.e. high T, isothermal gamma, discrepant \Gamma_HI or all of above. 
%%TH I mean all 3 parameters, revised
The inferred IGM thermal and ionization state [$\log T_0$, $\gamma$, $\log \Gamma_{\mathHI{}}$] is a manifestation of the aforementioned $b$-parameter distribution discrepancy, but now expressed as a quantitative measurement, which fully accounts for the parameter degeneracies with $\gamma$ and $\Gamma_{\mathHI{}}$. These results will be further discussed in the following section. 

%% JFH2 The color bar has no tick marks on it in these or the other similar plots, so I'm not sure what it is supposed to indicate???
\begin{figure*}
 \centering
     \includegraphics[width=1.0\textwidth]{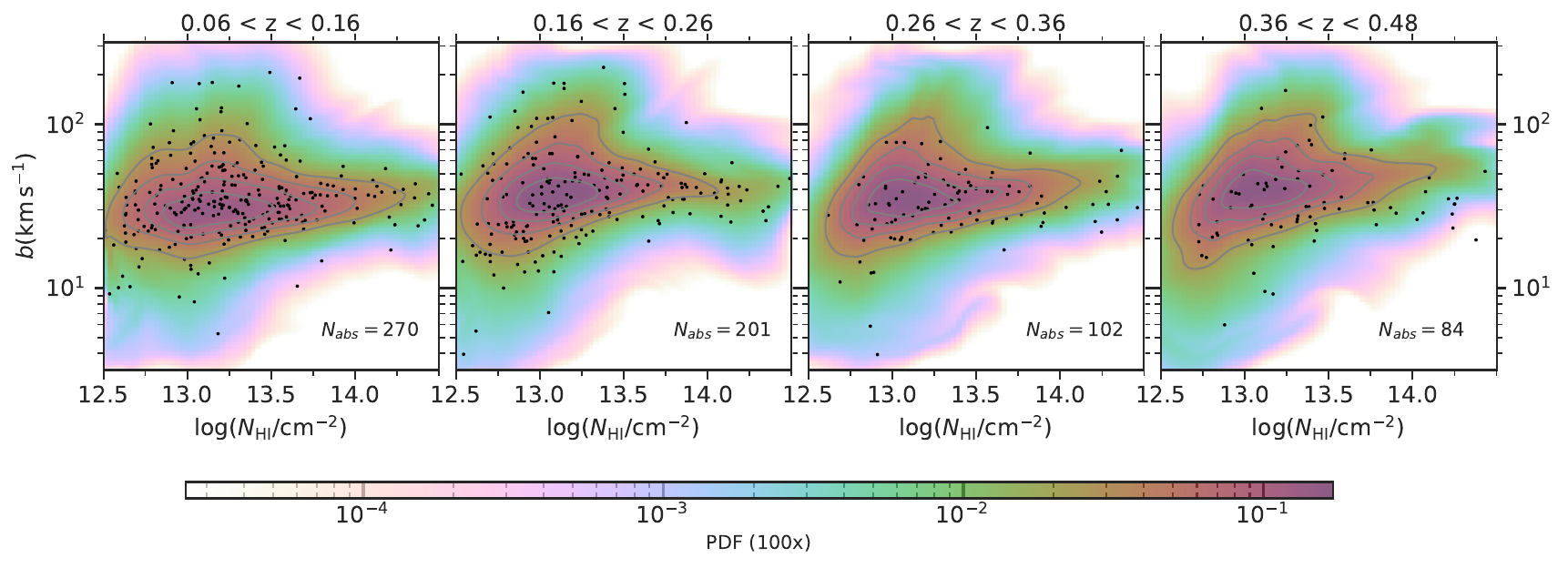}
  \caption{  Joint \bndist{}s emulated by our DELFI emulator based on the median values of the marginalized MCMC posterior at $z=$ 0.1, 0.2, 0.3 and 0.4. Black dots are the \bn{} data. The likelihood contours corresponding to 80,60,40, and 20 cumulative percentiles CDF are plotted as gray solid lines. For illustration purposes, the values of the pdf are multiplied by 100 in the colour bar. }
  \label{fig:fit_evol_z01}
\end{figure*} 

Figure~\ref{fig:fit_evol_z01} presents the \bn{} obtained from COS spectra alongside the corresponding \bndist{}s generated by our DELFI emulator. The model prediction is evaluated using the median values of the marginalized MCMC posterior. Grey solid lines indicate the 20th, 40th, 60th, and 80th percentiles of the cumulative distribution function (CDF).
%% JFH don't call these likelihood contours, since the likelihood is not a normalized probability distribution. Instead call it a prbobaility distribution or better yet, use the notation P(b,N | T_0, gamma). 
%%TH revised
%% JFH2 Cumulative percentiles is a bit confusing terminology, you should say probability density. You still didn't indicate what you plot here, i.e. median model. 
%%TH revised
%% JFH You need to clearly state which model you are showing for these b-N pdfs, i.e. that is the maxL model or the median model. The maxL is maybe preferred here, since the 1d marginals need not be close to the highest probability region of the posterior.
%%TH we are showing the median model, but it is very close to the maxL model. 
% VK grey dash colors are not very apparent neither are the dashes or thickness of lines - maybe try to modify figure 9 so that contours can be easily seen - you might want to try 25, 50 75 percentile if one can not see 4 lines
These plots demonstrate good agreement between the observational data and the \bndist{}s emulated by DELFI, even though the \bn{} sample size is relatively small, particularly in the $z=0.4$ bin.
%%JFH problem with englisht "while the precision is satisfactory" I also don't understand what you mean. Maybe you mean by eye it appears to be a good fit. 
%%TH I want to state that the data is mlimited but still the fit looks fine. revised

\begin{figure*}
 \centering
     \includegraphics[width=1.0\textwidth]{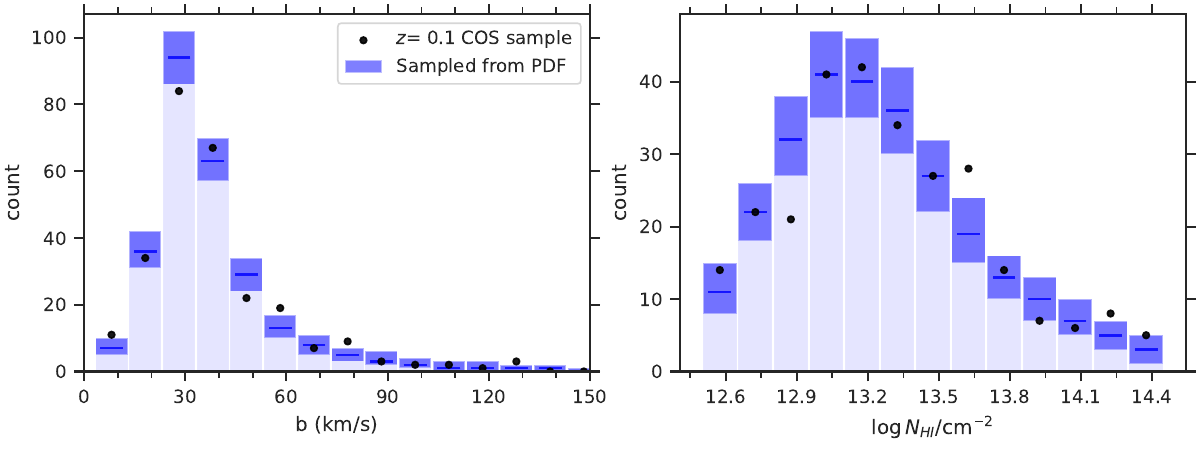}
  \caption{ The marginalized 1D $b$ and $N_{\mathHI}$ distributions of the data sample at $z=0.1$ compared with 5000 mock datasets with the same size, sampled from the \bndist{}s emulated based on the median values of the MCMC posteriors. The black dots represent our \bn{} data, and the blue bars indicate the mean value of the number of lines that fall in each bin for the 5000 datasets, whereas the blue shaded regions represent the 16th-84th percentiles calculated from the 5000 datasets. }
  \label{fig:bN_1d_z01}
\end{figure*}

To further evaluate the reliability of our inference results, we plot the marginalized 1D distributions of $b$ and $N_{\mathHI}$ for our sample at the $z=0.1$ bin in Fig.~\ref{fig:bN_1d_z01}. We compare the 1D marginalized $b$ distributions to these from 5000 mock datasets of the same size (different mocks forward-modelled based on the same observational data set), 
%% JFH by of the same size, do you mean that you forced them to have the same number of lines. Make this more clear. 
%%TH revised
each sampled from the \bndist{}s emulated at the median values of the MCMC posteriors. 
%% JFH Rather than generating all of the mocks at the median of the posterior, should you not be using maxL. This also applies to figure 9. 
The blue bars represent the median number of lines per bin across the 5000 datasets, while the blue shaded areas denote the 16th, 50th, and 84th percentiles 
%% JFH rather than say mean and 1\sigma. State that you are using 16 and 84th percentiles and the median. I would also not use the mean but the median. It won't change the answer much. 
%TH yes it is actually median
derived from these datasets. The results clearly demonstrate that our inference method effectively captures the marginalized 1D distributions of \bn{}. Due to the limited dataset size, there are fluctuations in the results, which are reflected by the 1-$\sigma$ error bars in the marginalized 1D distributions for both $b$ and $N_{\mathHI}$.

%% JFH I think you need to extend the \Gamma_HI grids in cases where you are hitting the grid boundary, namely Figure 5 at the low end, Figure 6 at the high-end (why the weird shape?) and Figure 7 at the high-end. 
% VK I think the same

\subsection{Evolution of the Thermal State of the IGM}
\label{sec:evol_z01}

 \begin{figure*}
\centering
    \includegraphics[width=0.95\linewidth]{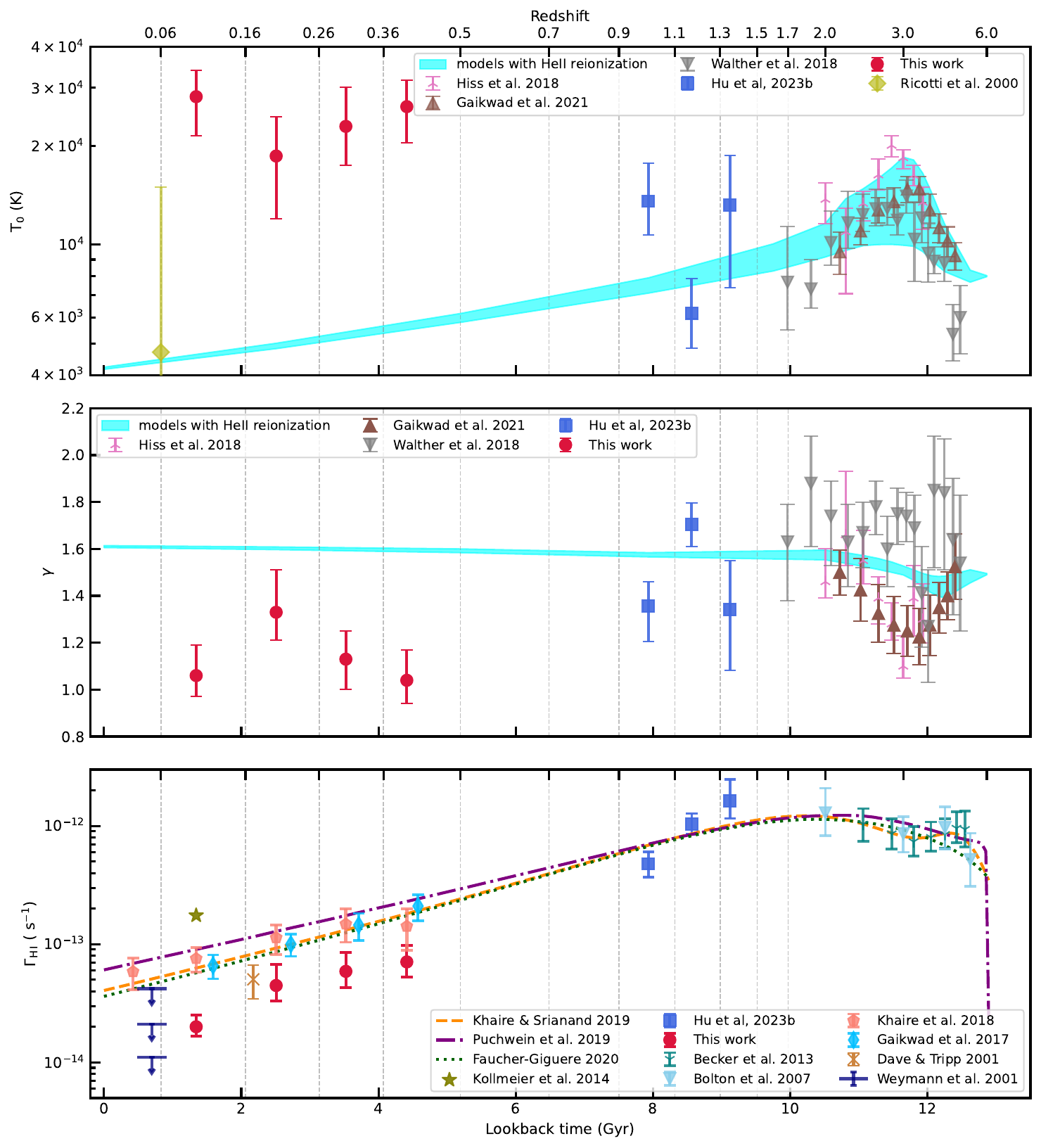}
  \caption{ Evolution history of $T_0$ (top), $\gamma$(middle) and $\log \Gamma_{\mathHI{}}$ (bottom) based on our inference results obtained from the COS data. Our results are shown as red dots, while measurements from other studies are displayed in different symbols and colors.
  The error bars stand for the 1-$\sigma$ error.
  The blue-shaded region in the top panel represents the range spanned by $T_0$ from hydrodynamical simulations of a large family of different HeII reionization models. The mock measurements based on Nyx simulation are shown in blue.}
  \label{fig:Thermal_evol_z01}
\end{figure*}

We summarize the evolution of $T_0$, $\gamma$, and $\Gamma_{\mathHI}$ across four redshift bins in Fig.~\ref{fig:Thermal_evol_z01}, in which we also present results from previous studies at higher redshifts \citep{Hu2023b, Hiss2018, WaltherM2019, Gaikwad2021}. Our measurements, along with their 1-$
\sigma$ errors, are shown as filled red data points with error bars. 
%% JFH you keep saying 1sigm auncertainties everywhere, but mean and standard deviation are not the same thing as median and 16-84th percentiles. Please be more clear with your language everywhere and state exactly what you show, which are 16-50-84 percentiles. 
%%TH revised
As a reference for current theoretical models, 
% VK It's not just cosmology -- maybe remove this preamble
%%TH revised
we plot the IGM thermal histories permitted by different Helium reionization models \citep{Onorbe2017, Onorbe2017b} as the cyan-shaded region.

%% JFH Did you extrapolate the cyan to low-z. How did you get this, I thought we didn't have it? Maybe another sentence should be added on the origin of the cyan. 
%%TH we have to z=0.2, But I will double if those are extrapolated

Our measurements indicate a significant discrepancy in $T_0$ at $z<0.5$, where our observed $T_0 = 28183$K, with $\sigma_{T_0} =6252$K
%% JFH just quot the value and its uncertainty rather than \sim 30k
%%TH revised
at $z =0.1$ is substantially higher than the values predicted by cosmological simulations ($T_0 \sim$4000K).
% VK predicted by cosmological simulations is vague -- say standard models of IGM and somewhere in before at appropriate location mention we refer standard models of IGM where the standard heating by UVB takes place
%% JFH say \sim 4000-5000 K or something here to indicate what the simulations predict. 
%%TH revised
We also notice that these higher-than-expected IGM temperatures, which exhibit an increasing trend towards lower redshifts, align with \citet[][blue data points in Fig.~\ref{fig:Thermal_evol_z01}]{Hu2023b}, which suggests an IGM $T_0 \sim$ 13,500 K at $z=1.0$. 
If the IGM is indeed much hotter than expected, such high IGM temperatures require the existence of additional heating sources not accounted for in current IGM models, particularly relevant around $z \sim 1.0$. Further discussion of this unexpected high temperature can be found in \S~\ref{sec:T0_discrepancy_z01}.

In addition, our measurements indicate that $\gamma$ is significantly lower than expected at low-$z$, with $\gamma \sim 1.0$ at $z=0.1$, suggesting that the low-$z$ IGM may be nearly isothermal, which might put important constraints on the aforementioned heating mechanism that caused the observed extremely high IGM temperature.  
However, due to the uncertainties in $\gamma$ measurements and the known degeneracy between $T_0$ and $\gamma$ \citep[see][]{Hu2022}, it remains uncertain whether the IGM is truly isothermal at low redshifts.
% VK this seems off, maybe add a line at the end saying clearly that one needs more precise measurements of gamma to confirm the isothermal nature seen here albeit with larger uncertainty. The point is for sure we are away from standard prediction but whether gamma is 1 or not is what you are talking about. It should be made more clear

%% JFH I think all the main words in the section title should be capital, even in British english, but I could be wrong. please check. 
%%TH revised
\subsection{Evolution of the H~{\sc i} Photoionization Rate and UVB}
\label{sec:evol_Gamma_z01}

Our measurements also provide insights into the $\Gamma_{\mathHI}$ evolution at $z<0.5$. 
In the bottom panel of Fig.~\ref{fig:Thermal_evol_z01}, we display our $\Gamma_{\mathHI}$ measurements across four redshift bins compared with results from previous studies \citep{Dave&Tripp2001, Bolton2007, Becker&Bolton2013, Kollmeier2014, Gaikwad2017, Khaire2019, Hu2023b}. 
We report $\Gamma_{\mathHI} = {-13.70}^{+0.10}_{-0.08}$, ${-13.35}^{+0.18}_{-0.13}$, ${-13.23}^{+0.16}_{-0.14}$, and ${-13.15}^{+0.14}_{-0.13}$ at $z=0.1$, $0.2$, $0.3$, and $0.4$, respectively. 
These values are noticeably lower than the predictions of the current UVB models such as the one presented in \citet{Khaire_Srianand2019} 
%% VK2 changes adding below
as well as \citet{Puchwein19} and \citet{FG2020}. 

While the \citet{Khaire_Srianand2019} model aligns well with other low-$z$ measurements derived from the \lya{} power spectrum \citep{Gaikwad2017b, Khaire2019} using the \citetalias{Danforth2016} low-$z$ \lyaf{} spectra, it is crucial to note that these measurements do not fully account for the degeneracy between the ionization and thermal state of the IGM. 
In their analyses, $\Gamma_{\mathHI}$ is measured using cosmological simulations with a fixed standard thermal history (specifically $T_0 \sim 5000$~K and $\gamma \sim 1.6$ at $z=0.1$). 
However, both higher IGM temperatures and higher $\Gamma_{\mathHI}$ reduce the optical depth, leading to increased transmission. 
Consequently, if the IGM is indeed hotter than assumed in standard models, the $\Gamma_{\mathHI}$ required to match the observed power spectrum must be lower. 
Our analysis breaks this degeneracy, favoring a lower $\Gamma_{\mathHI}$.

Finally, our new $\Gamma_{\mathHI}$ results, being lower than the theoretical model, are easily reconciled with the upper limits from \citet{Weymann2001} (see Fig.~\ref{fig:Thermal_evol_z01}). 
Given that the \citet{Weymann2001} study was conducted over two decades ago, we expect our findings to motivate a revisit of the topic using newer instrumentation. Newer observations could provide tighter constraints that would be critical for verifying the lower photoionization rates favored by our analysis.
%% JFH2 The Weymann label on Figure 11 says 2017, but it should say 2001. I believe there is a typo in the figure. 
%%TH fixed 

\section{Discussion}
\label{sec:discussion_z01}

\subsection{The Discrepancy in $T_0$ and $\gamma$}
\label{sec:T0_discrepancy_z01}

Many previous studies of the low-$z$ \lyaf{} have pointed out that the observed $b$-parameter distribution
%% JFH observed b-parameter distribution
%TH: revised
significantly surpasses the predicted value based on various simulations \citep{Gaikwad2017, Viel2017, Nasir2017, Bolton2022}.
Quantitatively, \citet{Viel2017} compares the marginalized $b$ distribution with various simulations, 
showing that the $b$ distribution at $z \sim$ 0.1 can be best recovered by the hydrodynamic simulations \citep[\texttt{P-GADGET-3}, see][]{Springel2005} with $ T_0 \gtrsim 10000$ K, 
while the theoretical model dictates that the $ T_0 \sim 5000$ at $z=0.1$.

\begin{figure*}
 \centering
     \includegraphics[width=1.0\textwidth]{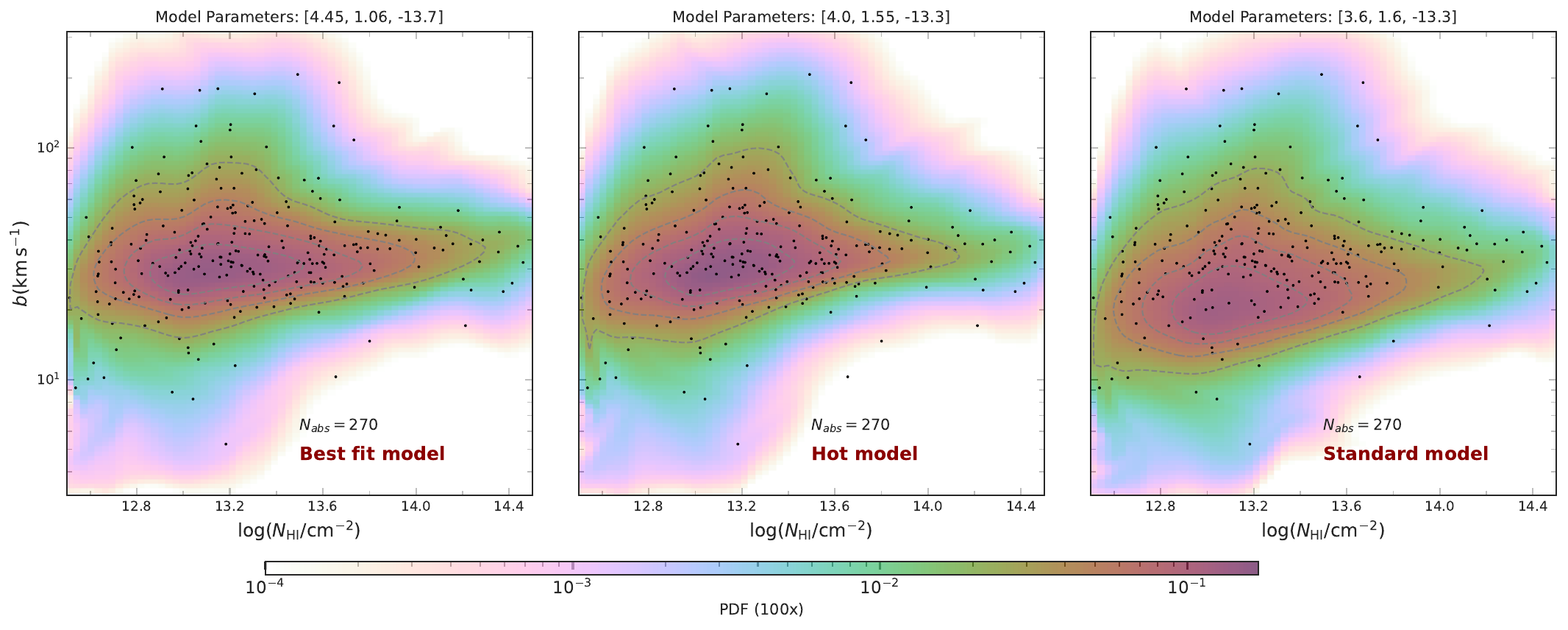}
  \caption{  Observed \bn{} data (black dots) at $z=0.1$ and joint \bndist{}s (color maps) at $z=0.1$ for three different thermal history models emulated by our DELFI emulator. The panels display: \textbf{(Left)} the ''Best fit model'' corresponding to the median values of our marginalized MCMC posterior ($[\log T_0, \gamma, \log (\Gamma_{\mathrm{HI}}/\mathrm{s}^{-1})] = [4.45, 1.06, -13.70]$); \textbf{(Middle)} a ''Hot model'' ($[4.00, 1.55, -13.30]$), which represents a hot model (but not isothermal) favoured by previous studies; and \textbf{(Right)} a ''Standard model'' ($[3.60, 1.60, -13.30]$ predicted by the theoretical model. %Joint \bndist{}s emulated by our DELFI emulator based on the median values of the marginalized MCMC posterior at $z=$ 0.1 vs the one recovered from [$\log T_0$, $\gamma$, $\log (\Gamma_{\mathHI}/s^{-1})$] = [4.0,1.55,-13.3], which represent model favoured by previous study based solely on the $b$ distribution. Black dots are the \bn{} data. The likelihood contours corresponding to 80,60,40, and 20 cumulative percentiles CDF are plotted as gray solid lines. 
  }
  \label{fig:bN_compare}
\end{figure*}

%% JFH2 Here in Figure 12 I think you should show the expected theoeretical value of 4000 km/s, rather than showing 10^4 K, which is the not very convincing inferred value from Viel. It makes the bigger case for your work, and we don't really care about the Viel temperature. 
%%TH maybe I can add one more plot for the standard model
Our results suggest the low-$z$ IGM is a hotter and more isothermal, with $T_0 \sim 30000$ K and $\gamma \sim 1.0$ at $z=0.1$. Such a result is largely consistent across four different redshift bins. To investigate this further, in Fig.~\ref{fig:bN_compare} we plot the \bndist{} recovered from our inference results ("Best fit model") and compare the "Hot model" ($[\log T_0, \gamma, \log (\Gamma_{\mathrm{HI}}/\mathrm{s}^{-1})] = [4.00, 1.55, -13.30]$), which represents a model favoured by previous studies based only on the $b$ distribution, and the "Standard model" ($[3.60, 1.60, -13.30]$) predicted by theoretical IGM evolution models. While both the "Best fit model" and "Hot models" reproduce the overall distribution of the observational data, the "Best fit model" provides a better match at the high $N_{\mathrm{HI}}$ end. As for the "Standard model" predicted by the IGM evolution model, the observed data points lie significantly above the predicted \bndist{} of the "Standard model", reaffirming that the canonical IGM evolution model significantly underestimates the line widths of the Ly$\alpha$ forest at $z < 0.5$.

As mentioned in \S\ref{sec:result_z01}, the unexpected thermal state of the IGM at low redshifts may be attributed to an unknown heating mechanism. It is also plausible that the discrepancies observed at $z \sim 0.1$ and $z \sim 1$ originate from the same heating source. 
Thus, our measurements combined with \citet{Hu2023b} indicate
% VK instead of hypothesize, we should say our measurements combined with Hu 2025 work indicate that 
%%TH revised
that this heating mechanism becomes significant around $z \sim 1$ and persists down to $z=0$. 
If this hypothesis holds true, it would dramatically alter our understanding of \ac{IGM} physics, 
highlighting the urgent need to explore possible sources such as dark matter heating \citep{Araya2014, Bolton2022DarkPhoton}, gamma-ray heating \citep{Puchwein2012}, or feedback processes from galaxy formation that remain poorly constrained at low redshifts \citep[see][]{Springel2005, Croton2006, Sijacki2007, Hopkins2008, Christiansen2020, Tillman2023, Tillman2023b,Khaire2023,Khaire2023_halos,Hu2023}. Another intriguing candidate is dust heating \citep{Inoue2010,Bolton2022}, whose heating rate scales as $u_\text{dust} \propto \Delta^{(1+\gamma)/6}$. 
%% JFH2 Not clear to me why dust heating lowers the value of gamma? I'm not following your reasoning here. 
%%TH I think you are right. revised it 
Observational evidence suggests that dust may be more abundant in the IGM than previously assumed \citep{menard2010}, and recent simulations demonstrate that dust grains can survive galactic winds and be transported into the IGM \citep{Chen2024}.
% VK Bolton is not annihilation, so remove annihilation say non-standard dark matter 
%%TH revisef
%% JFH What about dust heating, why are you not mentioning that in these discussions. 
%%TH  added

\subsection{The Impact of Turbulence}
\label{sec:Turbulence}

An alternative explanation of the observed higher-than-expected $b$-parameter is the existence of small-scale turbulence in the low-$z$ IGM, which increases the width of the observed \lya{} lines \citep{Nasir2017,Viel2017,Gaikwad2017,Bolton2022}. In this section, we assess such a hypothesis quantitatively by applying our inference method to the COS \lyaf{} dataset at $z \leq  0.5$, with standard thermal history and flexible small scale IGM turbulence.

\begin{figure*}
 \centering
    \includegraphics[width=0.49\textwidth]{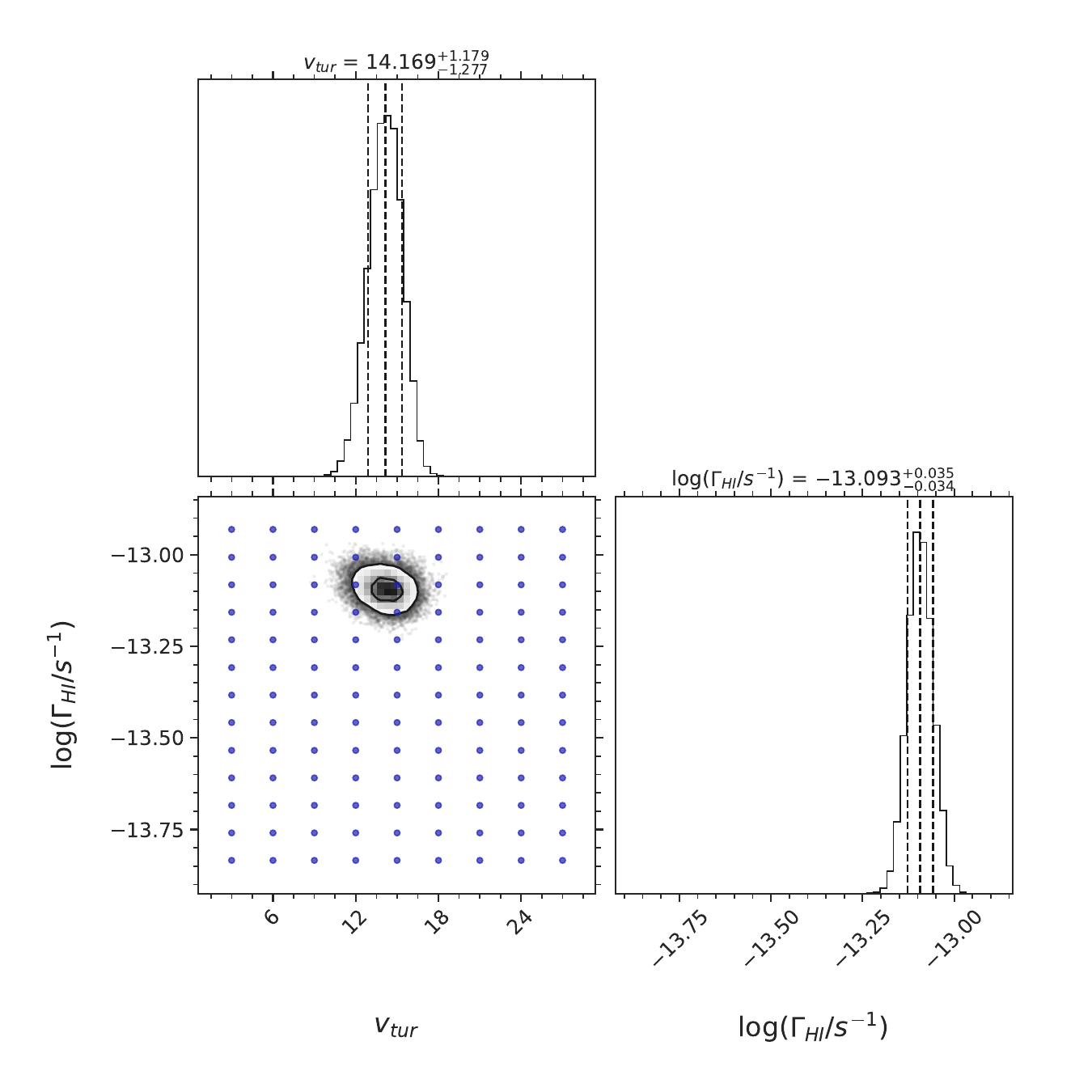}
    \includegraphics[width=0.49\textwidth]{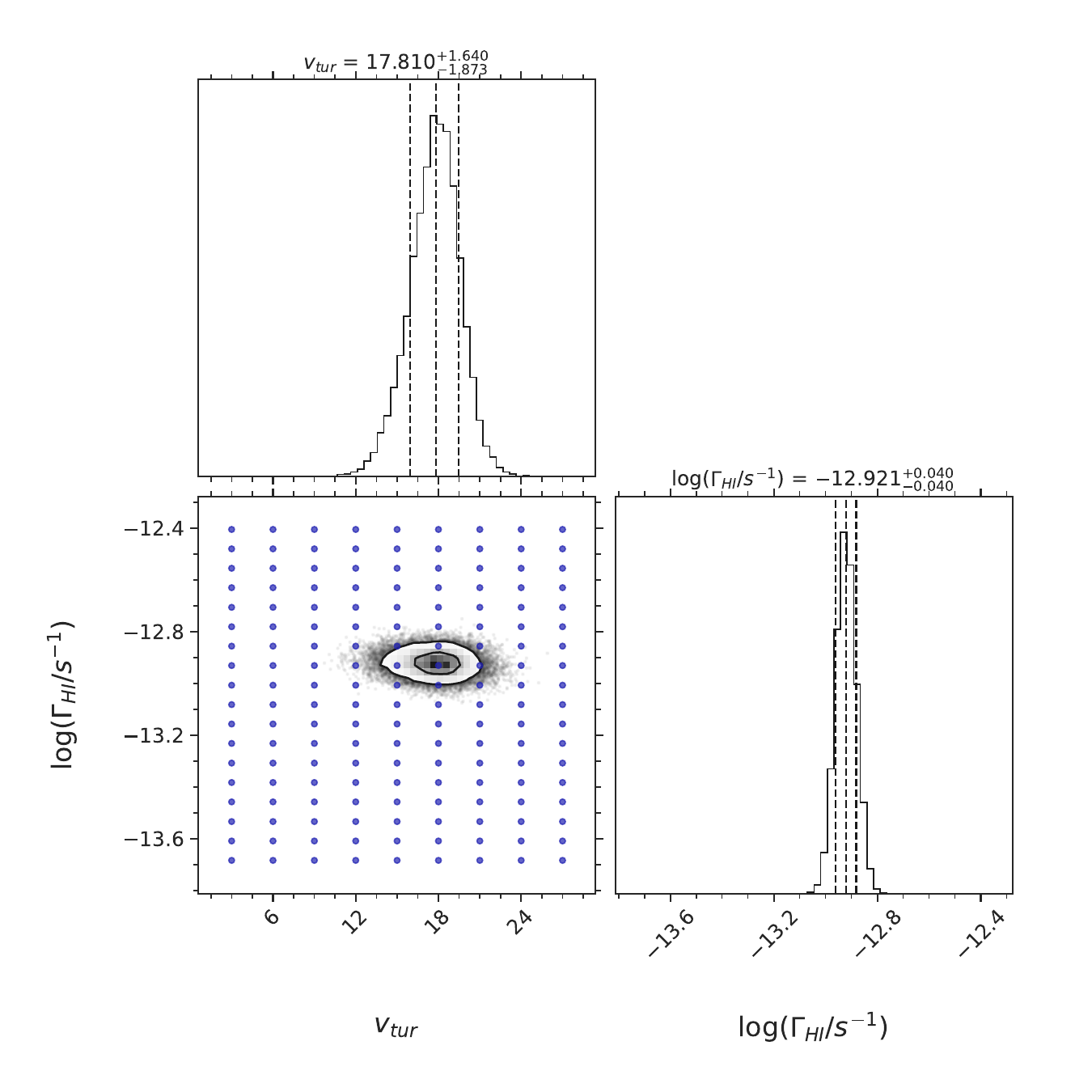}
    \includegraphics[width=0.49\textwidth]{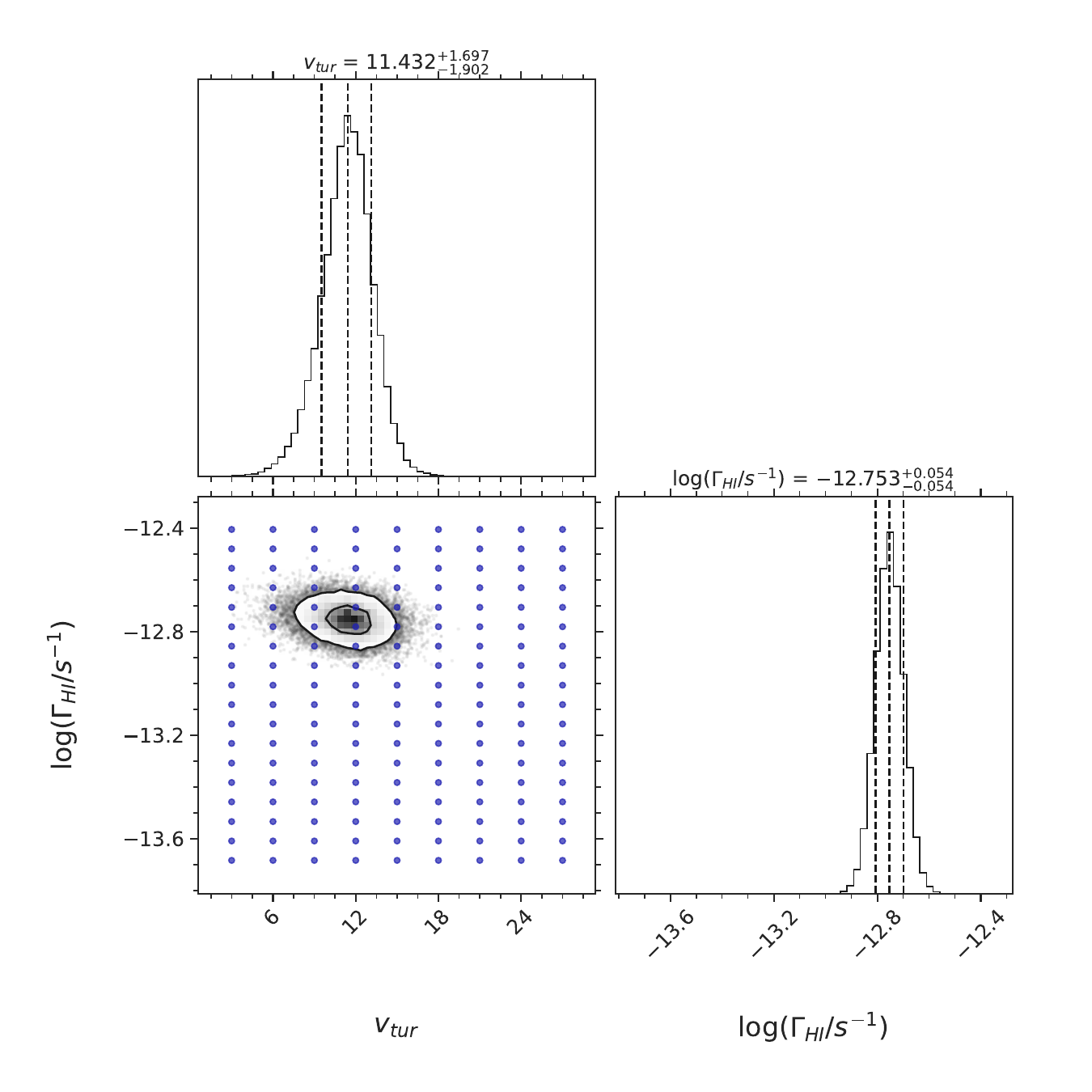}
    \includegraphics[width=0.49\textwidth]{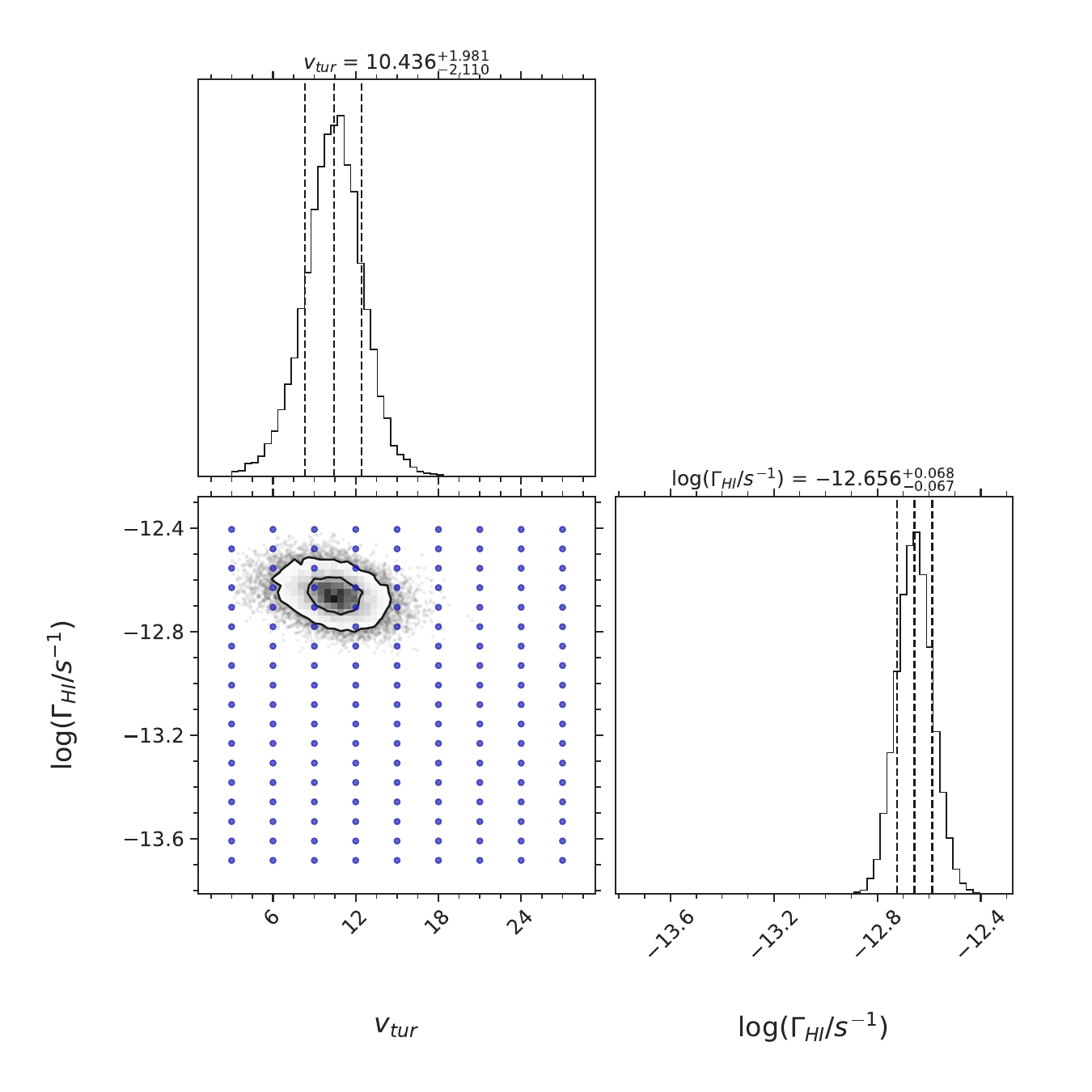}
  \caption{ Posteriors obtained by applying our inference method on the $v_\text{tur}$-$\Gamma_\mathHI$ grid at $z=0.1,0.2,0.3$ and 0.4. Projections of the parameter grid used for generating models are shown as blue dots. The inner (outer) black contour represents the projected 2D 1(2)-sigma interval. The dashed black lines indicate the 16, 50, and 84 percentile values of the marginalized 1D posterior. }
  \label{fig:corner_tur}
\end{figure*}

\begin{figure*}
 \centering
    \includegraphics[width=0.99\textwidth]{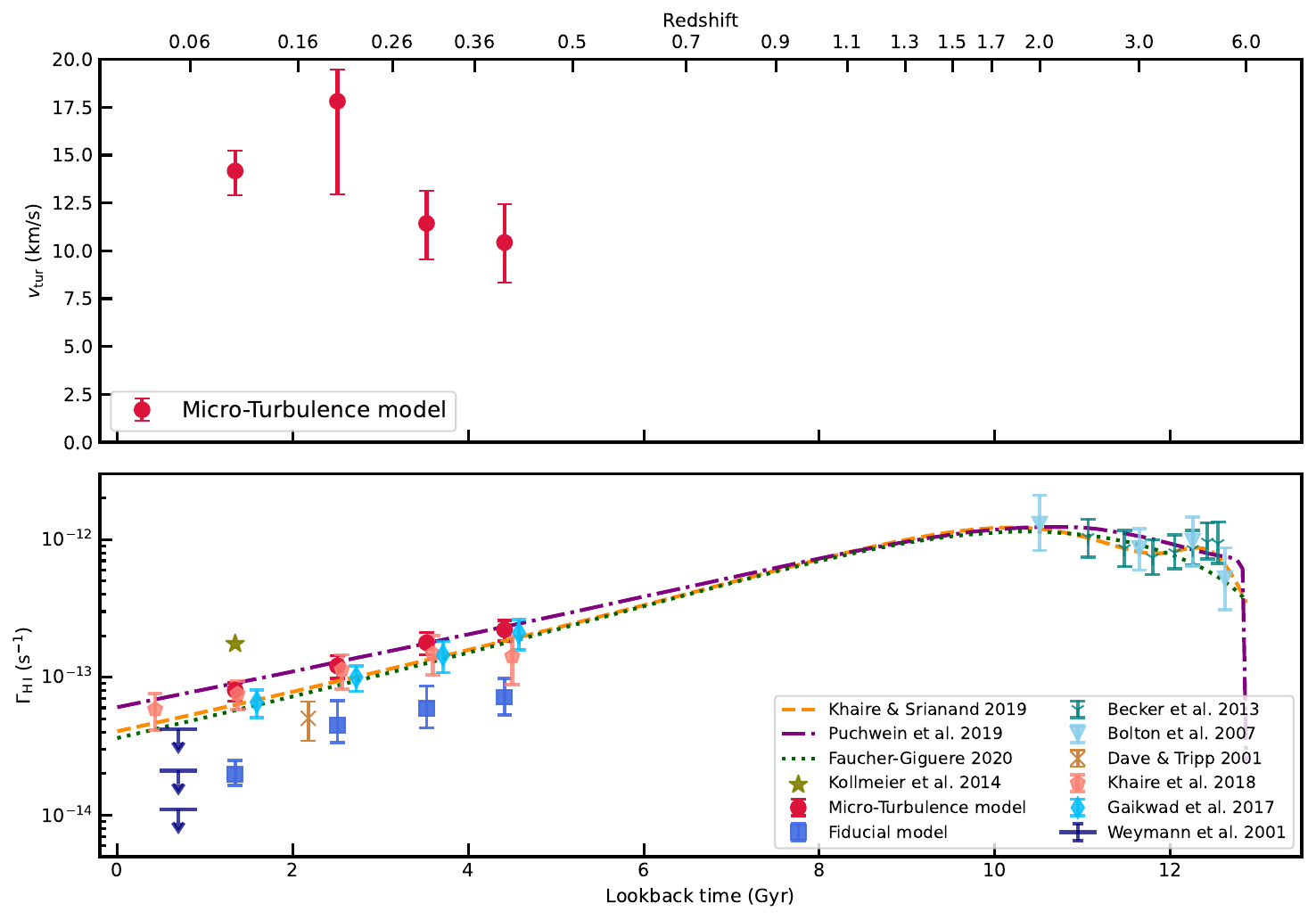}
  \caption{ Evolution history of $v_\text{tur}$ (top) and $\log \Gamma_{\mathHI{}}$ (bottom) based on standard thermal model and altered small-scale velocity. The results are shown as dark red dots, while measurements from other studies are displayed in different colours. The error bars stand for the 1-$\sigma$ error.}
  \label{fig:evol_tur}
\end{figure*}

In practice, we model the small turbulence in the IGM by adding a Gaussian component {$N(0,v_\text{tur})$} 
%% VK2 I have replaced the old notation \sigma_{v} to v_tur everywehre. This was confusing a lot. We clearly say how we incorporate it below and need not to use different notations. 
to the peculiar velocity along line-of-sight, where $v_\text{tur}$ is the standard deviation in km/s.
This random velocity component
is added to each simulation cell with 
%% JFH2 Is this comoving or proper, need to state explicitly!
$\Delta L$= 2.4 kpc/$h$ (comoving).
%% JFH Quoute the value in kpc/h not Mpc/h. Acutally I would probably get rid of the 1/h and quote in ckpc. Also make it clear somewhere that you mean comoving throughout the text or use ckpc or cMpc. 
%%TH checking it
To quantitatively constrain the turbulence, we post-process the simulation and generate skewers with $v_\text{tur}$ = 3-27 km/s, in steps of 3 km/s.
%% JFH this looks awkard. Say 3-27 in steps of 3 or something like that. 
%TH revised
To investigate turbulence as an alternative explanation for the higher-than-expected IGM temperature, we adopt the standard Nyx model with $[T_0, \gamma] \sim [4000~\mathrm{K}, 1.6]$ at $z=0$. These values are insensitive to He~II reionization models because the thermal evolution is dominated by the expansion of the Universe, forcing the thermal states of different models to converge at low $z$ (see Fig.~\ref{fig:Thermal_evol_z01}). 
%% JFH2 Explain why the standard model is standard and what you mean by that. In other words, cite the converging temperature evolution for HeII reionization models in Figure 11 upper panel. 
%%TH revised
For each  $v_\text{tur}$ value, we generate forward-modelled mock spectra with 13 different UVB
photoionization rate, $\Gamma_{\mathHI}$, following the prescription given in \S\ref{sec:simulations}. We then apply our inference framework on the $v_\text{tur}$-$\Gamma_{\mathHI}$ parameter gird following the procedure discussed in \S\ref{sec:inference}. To monitor the evolution of such turbulence, we conduct the inference at all four redshift bins individually, and obtain that 
%$v_\text{tur}$ = 14,17,12,11 km/s 
$v_\text{tur}$ = 14,18,11,10 km/s %% VK2 change
for $z= 0.1,0.2,0.3$ and 0.4; while the corresponding $\log (\Gamma_{\mathHI}/s^{-1}) =-13.1, -12.9,-12.8$, and -12.7. Such a $v_\text{tur}$ is consistent with
with the one derived in \citet{Bolton2022}, which yield $v_\text{tur} \sim$ 15 km/s at $z=0.1$. 
% VK I don't think Bolton paper had does such a rigorous analysis, you can mention how they arrived in to this number in a line and say that our numbers are consistent with it
The inference results are shown in Fig.~\ref{fig:corner_tur}, and the evolution history of $v_\text{tur}$ and $\log \Gamma_{\mathHI{}}$ are shown in Fig.~\ref{fig:evol_tur}. Interestingly, it can be seen that the $v_\text{tur}$ required to match the observation increases toward low-$z$, suggesting that the discrepancy between the observation and simulation in $b$-parameter must be caused by continuous sources that increase 
toward low-$z$.

% VK in this section we need a plot that @Teng had in his thesis presentation that shows altering the peculiar velocities do not change the whole velocity distribution -- this is to justify that we are not changing anything hugely -- a tiny change in the peculiar velocity distribution works.

In addition, we notice that with the standard thermal model and altered small-scale velocity, our inference method suggests higher $\Gamma_{\mathHI}$ values. 
% VK no these are not higher -- they are consistent with models and other measurements. You can say that our measurements are lower and not you get consistent with models and other measurements. Point to make is once you match b distribution with non-thermal motions then you get back the standard UVB
This is mainly because, while the $v_\text{tur}$ has no noticeable impact on the \dndz{}, both $T_0$ and $\Gamma_{\mathHI}$ have similar correlation on the \dndz{}, i.e., both higher $T_0$ and $\Gamma_{\mathHI}$ suppress the formation of the \HI{} absorbers in the IGM, causing degeneracy in the inference results.

%% JFH Fiducial model is a weird label here on the plot in th elower panel. Also I suggest you change the color scheme and  always show your measurements in red. 
%%TH i will call it standard then. TH making new plot for it

\subsection{ The Impact of the Resolution}
\label{sec:discuss_Res}

Although our measured $T_0$ and $\gamma$ deviate strongly from theoretical expectations, our estimates of $\Gamma_{\mathHI}$ are in good agreement with the \citet{Dave&Tripp2001} results based on \ac{HST}/\ac{STIS} data. 
However, that study also reports that the $b$-parameter distribution of the low-$z$ \lyaf{} is consistent with simulations. 
While their simulations included sub-grid prescriptions for feedback, which might affect line broadening, recent work by \citet{Hu2023} suggests that even extreme feedback does not significantly alter the \bndist{} of the IGM. 
This leaves open the possibility, however unlikely, that the discrepancy we observe in the $b$-parameter distribution arises from an overestimation of the COS resolution. 
If the true COS resolution is lower than the quoted value, our forward models would underestimate the instrumental line broadening, resulting in simulated \lya{} lines that are systematically narrower than those observed. 
In this section, we test whether the observed discrepancy in the $b$-parameter distribution can be attributed to such an overestimation of the COS resolution.

%% JFH You need to add more sentences to make this more clear. In other words explain the basic physical effect and its direction first. In other words, if the COS resolution is actually lower than the quoted value, then we are not broadening the lines enough in our forward models, and this difference in LSF broadening could be causing the the lines to be broader in the observed data than in the forward models, or something like that. You need to spell out exactly how this sysetmatic would cause you to infer higher than epxected temperatures. 
%%TH revised
Firstly, we aim to determine the required resolution of HST COS to observe the aforementioned discrepancy if the incorrect COS LSF is the sole factor, i.e., if the IGM is neither hotter than expected nor affected by additional turbulence.

Similar to the analysis in \S\ref{sec:Turbulence}, we apply our inference method to a 2D parameter grid consisting of spectral resolution and the UV background $\Gamma_{\mathHI}$. In this test, we assume that the “true” COS \ac{LSF} is unknown and seek to determine what \ac{LSF} would be required to reproduce the observed $b$-parameter distribution at $z=0.1$. To this end, we generate forward-modeled mock spectra using 10 different Gaussian \acp{LSF}, with resolution ranging from 10 to 100 km/s (in FWHM) in steps of 10 km/s. This procedure is applied to all standard Nyx models (model T000, with $T_0 \sim 4000$ K and $\gamma \sim 1.6$ at $z=0.1$) across the 13 different $\Gamma_{\mathHI}$ values described in \S\ref{sec:Gamma_grid}. To ensure comparability with the observations, our VP-fitting program still employs the tabulated COS \ac{LSF} for both the mock and observed spectra, as detailed in \S\ref{sec:vpfit_z01}.

\begin{figure}
 \centering
    \includegraphics[width=0.49\textwidth]{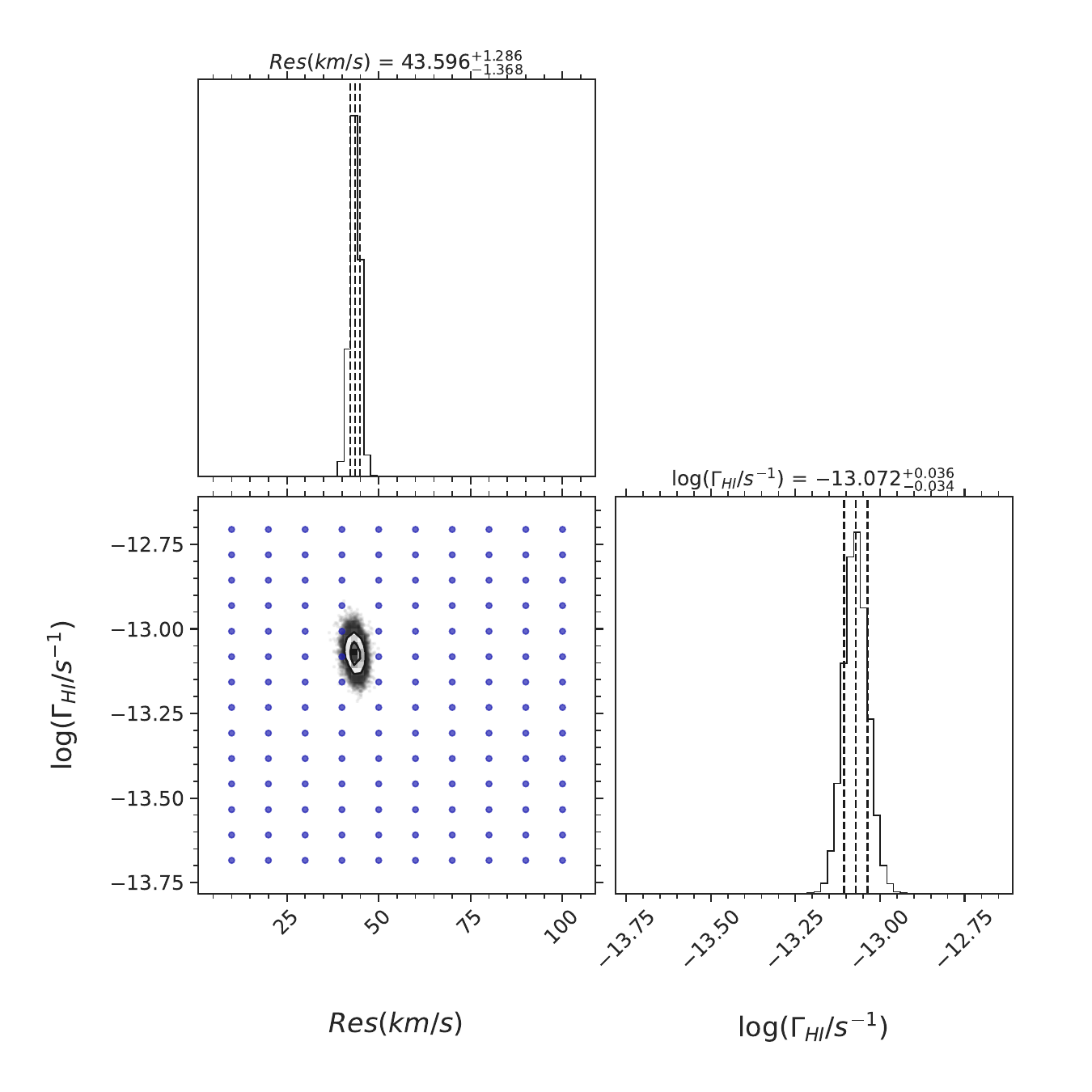}

  \caption{ Posteriors obtained by applying our inference method on the Resolution-$\Gamma_\mathHI$ grid at $z=0.1$. Projections of the parameter grid used for generating models are shown as blue dots. The inner (outer) black contour represents the projected 2D 1(2)-sigma interval. The dashed black lines indicate the 16, 50, and 84 percentile values of the marginalized 1D posterior. }
  \label{fig:corner_Res}
\end{figure}

We run our inference method to find the resolution required to obtain the observed \bndist{}, and the inference result is shown in Fig.~\ref{fig:corner_Res}.
Given the observed $b$ distribution peaking at $\sim 30$ km/s, we find that the required resolution (FWHM) is about $43.6$ km/s, which corresponds to a resolution $R \sim$ 7000, while the reported $R$ for HST COS is 15000 to 20000.%, making it very unlikely to be true. 

%% JFH Well but these models look to be a terrible fit to the 1d b-distribution. YOu need to make a bigger deal out of that, and how this makes this to be unlikely the solution to the problem. But I'm really not following why you set the resolution to always be the same when you ran VP-fit. 
%%TH because this is to reproduce the case where the tabluated COS LSF is wrong, but we observe the lyaf using the tabluated LSF, and we try to figure out how worng does the tabluated COS LSF has to be to reproduce the observation.

In \citet{2009COS_LSF}, the reported COS LSF is carefully examined using the COS spectra of the O9 Ib supergiant star Sk 155 in the Small Magellanic Cloud (V=12.4), 
% VK what's is V?
where the COS spectra are compared with those observed with STIS E140H spec (R$\sim$114000), which are then convolved with the modeled COS LSF. The close alignment between these spectra confirms the high accuracy of the reported COS LSF.

Here, we follow the aforementioned method and compare the COS spectra with the higher-resolution STIS E140M spectra, which cover 1144$\sim$1729 \AA{} and has a reported resolution of approximately 45000—roughly three times that of the COS. We use the STIS E140M as 'intrinsic' spectra and convolved them with the HST COS LSF as tabulated in {\tt linetools}. This allows us to compare the reported COS LSF with the actual COS spectra to determine if the reported HST COS resolution is accurate. 

\begin{figure*}
 \centering
     \includegraphics[width=1.0\textwidth]{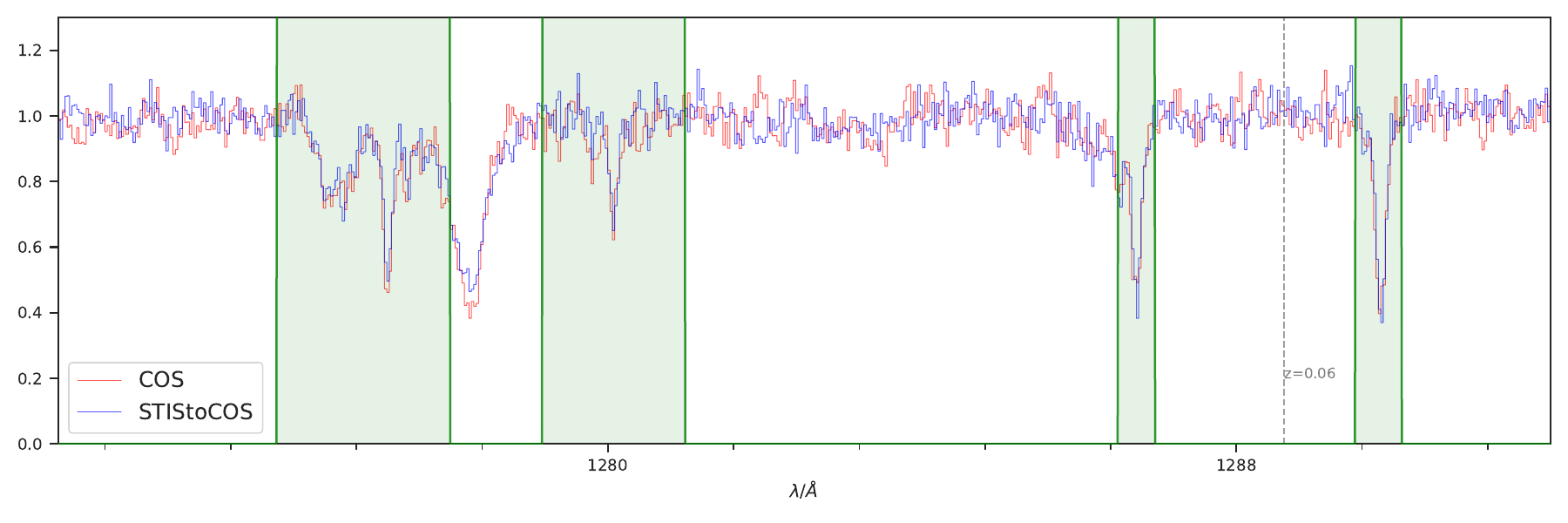}
  \caption{A segment of the COS G130M spectrum of PHL1811 (red) compared with the corresponding STIS E140M spectrum convolved COS G130M LSF (blue). The green shaded region indicates the masked region due to the existence of metal absorption lines.}
  \label{fig:spec_lsf_compare}
\end{figure*} 

To this end, we utilize STIS E140M spectra of the object PHL1811, PG1216+069, 3c273, and H1821+643 \citep{Jenkins2005, Tripp2008, Williger2010}, whose COS spectra are also examined in this work, and perform a detailed comparison. Visual inspection suggests that convolving these very high-resolution STIS E140H spectra with the COS LSF models leads to an excellent match to the observed COS FUV spectra. Fig.~\ref{fig:spec_lsf_compare} shows a segment of the COS G130M spectrum of PHL1811 (red) compared with the corresponding STIS E140M spectrum convolved COS G130M LSF (blue). The green shaded region indicates the masked region due to the existence of metal lines. It can be seen that, for these narrow metal lines, the high-resolution STIS spectrum convolved with reported COS LSF matches well with the observed COS spectrum, suggesting that the reported COS LSF model is accurate and reliable. We therefore conclude that the observed discrepancy in the $b$ parameter is not caused by the overestimation of the HST COS resolution.

\section{Summary and Conclusion}
\label{sec:conclusion_z01}

In this paper, we make use of 82 archival HST G130/G160 quasar spectra, 
from which we obtain the \bndist{} distribution and line density \dndz{} over the redshift range $0.06 < z < 0.48$ in four redshift bins.
We then measure the thermal and ionization state of the IGM following a
machine-learning-based inference method presented in \citet{Hu2022} for this redshift range for the first time.
% VK do we need to say ML based inference? Our inference is not using ML - only b-N is emulated by ML and inference is Bayesian mcmc. Right? 
%%TH that true, but i am wondering ML-based might be more interesting to ppl?
We summarize our results below:
\begin{itemize}

     \item We Voigt-profile fit the \lyaf{} in all 82 quasar spectra using a fully automated \vpfit{} wrapper and obtain \bn{} for 731 lines within the parameter limits.
     We use the metal identifications from the \citetalias{Danforth2016} to generate our metal masks, filtering out 74 contaminants besides \lya{} absorption lines, and obtain a final sample of 657 \lya{} lines across a total pathlength of $\Delta z=$4.43. 
     % VK To the causal reader it will seem like only 657 Lyman alpha lines are there in the dataset, but the NHI range that we choose decides it. 
     %%TH revised

     \item We employ the \citet{Hu2022} inference method, 
     which simultaneously measures $[T_0, \gamma, \Gamma_{\mathHI}]$ from the \bndist{} and \dndz{}, 
     with the help of neural density estimators and Gaussian process emulators trained on a suite of 51 Nyx simulations, each having a different IGM thermal history.
     It enables us to measure the IGM thermal and ionization state with high precision even with limited data.

     %% JFH2 Can you just cit the Table here?
     %%TH done 
     \item  We obtain $T_0 = {28183}^{+5700}_{-6804}$ K and $\gamma = {1.06}^{+0.13}_{-0.09}$ at $z=0.1$ (See Tab.\ref{tab_inf_result_z01} for other redshift bins). These measurements suggest that the IGM is actually much hotter than expected and close to isothermal at $z<0.5$, with
     %% VK2 change below to remove dangaling with
     $T_0$ roughly a factor of seven above the canonical prediction and $\gamma$ consistent with unity.
     
     %\item We compare our findings with previous work, which reports unexpected high $b$-parameters compared with various simulations based on observational data at $z \sim 0.1$. According to previous work, the observed high $b$ values, if caused by thermal broadening, implies a hotter-than-expected IGM. In this work, we analysis problem quantitatively and take into account the degeneracy between $T_0$, $\gamma$ and $\Gamma_{\mathHI}$. Benefited from the novel method and larger dataset, our resluts imply a $T_0 \sim $ 30000K and $\gamma \sim 1.0$ at $z =0.1$, significantly hotter than previous estimation where $T_0 \sim 10000$K.
     % VK this point is downplaying what we get. I would rewrite it to say the b-parmater was already hinting something, we resolve it but the implication is IGM is too hot. 
     %TH working on it.
     %%TH revised
     
    \item We compare our results with previous studies that reported unexpectedly large $b$-parameters at $z \sim 0.1$, exceeding those predicted by simulations. Previous works interpreted these broad \lya{} lines, if dominated by thermal broadening, as hint for a hotter-than-expected IGM. In this work, we analysis this problem quantitatively and take into account the degeneracy between $T_0$, $\gamma$ and $\Gamma_{\mathHI}$. Benefiting from a larger dataset and our machine-learning-based inference method, we find $T_0 \sim 3 \times 10^4$ K and $\gamma \sim 1.0$ at $z=0.1$, indicating a substantially hotter and more isothermal IGM than previous work, which suggest that $T_0 \sim 1 \times 10^4$ K at z $\sim 0.1$.

     \item We successfully measure the $\Gamma_{\mathHI}$ at four 
     %redshif %% VK2 correction
     redshift bins, reporting $\Gamma_{\mathHI}$=${-13.70}^{+0.10}_{-0.08}$, ${-13.35}^{+0.18}_{-0.13}$, ${-13.23}^{+0.16}_{-0.14}$,
     and ${-13.15}^{+0.14}_{-0.13}$  at $z=0.1, 0.2$, 0.3 and $0.4$ respectively. These measurements are noticeably lower than the predictions of the UVB model presented in \citet{Khaire_Srianand2019}, and the measurements of \citet{Gaikwad2017, Khaire2019} using the \lya{} power spectrum based on \citetalias{Danforth2016}, but agree with the measurements made by \citet{Dave&Tripp2001} based on the STIS data.
     However, it is worth mentioning that all previous measurements do not take the potential degeneracy between the IGM thermal and ionization state into account.

      \item Our results may point to an additional heating mechanism that becomes important around $z \sim 1$ and persists to the present. Such a process could explain the discrepancies seen at both $z \sim 0.1$ and $z \sim 1$, and would imply new physics shaping the IGM thermal state. Possible contributors include dark matter heating, gamma-ray heating, feedback from galaxy formation, or dust heating, which is more widespread in the IGM than previously assumed.
      %% JFH2 I would get rid of the "the latter of which can also reduce...."
        %%TH done
     % VK: this is a repeated line and you need not to cite these references again
     %% JFH mention dust heating. 
     %%TH revised

    %%% JFH clean up the language surrounding resolution and make it clear agian the direction of the effects, etc. See my previous comment. 
    %%TH  revised
     \item An alternative explanation of the observed higher-than-expected $b$ parameter is the existence of small-scale turbulence in the low-$z$ IGM, which increases the width of the observed \lya{} lines. To this end, we perform our inference method on a $v_\text{tur}$-$\Gamma_{\mathHI}$ grid to conclude that if the observed discrepancy is indeed caused by turbulence in small-scale, it would need velocity dispersion with $v_\text{tur} \sim 14$, 18, 11, and 10 km/s at $z=$ 0.1, 0.2, 0.3 and 0.4 respectively.  Furthermore, the increase in $v_\text{tur}$ towards low-$z$ implies that the discrepancy between observed and simulations in $b$-parameters, whether caused by turbulence, must be driven by continuous sources that intensify towards low-$z$.

    \item In addition, we evaluate whether the observed effect can be caused by overestimation of the COS resolution. We find that it requires an effective resolution of $43.6$ km/s 
    %% JFH2 earlier in the text you quoted \sim 40, now you say 43.5 make the way you quote this number consistent please. 
    ($R\sim $7000) to cause the observed \bndist{}. Furthermore, we perform a detailed comparison between the HST COS spectrum and the spectrum observed with HST STIS E140M, which has a much higher resolution, for four objects. The comparison suggests that the reported COS resolution (LSF) is reliable, and the observed discrepancy in the $b$-parameter could not be caused by the resolution effects solely.

\end{itemize}

%% JFH You need a few more sentences here on what ought to be done in the future etc. 
%%TH revised 
In addition to the HST/COS spectra analyzed in this work and the HST/STIS data presented in \citet{Hu2023b}, we are expecting more archival STIS spectra spanning $0.5 < z < 2$ \citep{Chen2023hst}. After careful metal line identification, these spectra can be incorporated into our framework to further extend the redshift coverage. %We are also in the process of obtaining additional HST observations. 
In the future, we plan to apply our inference methodology to simulations that include more sophisticated and diverse feedback prescriptions, such as those implemented in EAGLE \citep{EAGLE} and the CAMELS suite \citep{camels_presentation}. These future efforts will allow us to investigate potential heating mechanisms that may contribute to the unexpectedly hot low-$z$ IGM revealed by this work.

\begin{acronym}
	\acro{AGN}{active galactic nuclei}
	\acro{CMB}{Cosmic Microwave Background}
	\acro{COS}{Cosmic Origins Spectrograph}
	\acro{DELFI}{density-estimation likelihood-free inference}
	\acro{DM}{dark matter}
	\acro{DLA}{damped Ly$\alpha$}
	\acro{GP}{Gaussian process}
	\acro{HIRES}{High Resolution Echelle Spectrometer}
	\acro{HST}{Hubble Space Telescope}
	\acro{IGM}{intergalactic medium}
	\acro{KDE}{Kernel Density Estimation}
	\acro{KODIAQ}{Keck Observatory Database of Ionized Absorbers toward QSOs}
	\acro{LD}{least absolute deviation}
	\acro{LLS}{Lyman limit systems}
	\acro{LS}{least squares}
	\acro{LSF}{line spread function}
	\acro{MCMC}{Markov chain Monte Carlo}
	\acro{MW}{Milky Way}
	\acro{NDE}{neural density estimation}
	\acro{PCA}{principal component analysis}
	\acro{PDF}{probability density function}
	\acro{PKP}{\ac{PCA} decomposition of \ac{KDE} estimates of a \ac{PDF}}
	\acro{QSO}{quasi-stellar objects}
	\acro{SNR}{signal-to-noise ratio}
	\acro{STIS}{Space Telescope Imaging Spectrograph}
	\acro{TDR}{temperature-density relation}
	\acro{THERMAL}{Thermal History and Evolution in Reionization Models of Absorption Lines}
	\acro{UV}{ultraviolet}
	\acro{UVB}{ultraviolet background}
	\acro{UVES}{Ultraviolet and Visual Echelle Spectrograph}
	\acro{WHIM}{warm hot intergalactic medium}
        \acro{CASBaH}{COS Absorption Survey of Baryon Harbors}
\end{acronym}

\section*{Acknowledgements}

The authors thank the ENIGMA members\footnote{http://enigma.physics.ucsb.edu/} and Joe Burchett for helpful discussions and suggestions. 

Calculations presented in this paper used the hydra and draco clusters
of the Max Planck Computing and Data Facility (MPCDF, formerly known
as RZG). MPCDF is a competence center of the Max Planck Society
located in Garching (Germany).
This research also used resources of the National Energy Research Scientific Computing Center (NERSC), a U.S. Department of Energy Office of Science User Facility located at Lawrence Berkeley National Laboratory, operated under Contract No. DE-AC02-05CH11231 
In addition, we acknowledge PRACE for awarding us access to JUWELS hosted by JSC, Germany.
JO acknowledges support from grants BEAGAL18/00057 and CNS2022-135878 from the Spanish Ministerio de Ciencia y Tecnologia.

\section*{Data Availability}

The simulation data and analysis code underlying this article will be shared on reasonable request to the corresponding author. 
%%%%%%%%%%%%%%%%%%%%%%%%%%%%%%%%%%%%%%%%%%%%%%%%%%



\bibliographystyle{mnras}


\bibliography{references.bib}

%%%%%%%%%%%%%%%%%%%%%%%%%%%%%%%%%%%%%%%%%%%%%%%%%%

%%%%%%%%%%%%%%%%%%%%%%%%%%%%%%%%%%%%%%%%%%%%%%%%%%

% Don't change these lines
\bsp	% typesetting comment
\label{lastpage}
\end{document}